\documentclass[twocolumn,superscriptaddress,preprintnumbers,amsmath,amssymb,longbibliography]{revtex4-1}

\newcommand{\IMSS}{Muon Science Laboratory and Condensed Matter Research Center, Institute of Materials Structure Science, High Energy Accelerator Research Organization (KEK-IMSS), Tsukuba, Ibaraki 305-0801, Japan}
\newcommand{\Sokendai}{Department of Materials Structure Science, The Graduate University for Advanced Studies (Sokendai), Tsukuba, Ibaraki 305-0801, Japan}
\newcommand{\MCES}{Materials Research Center for Element Strategy, Tokyo Institute of Technology (MCES), Yokohama, Kanagawa 226-8503, Japan}

\usepackage[dvipdfmx]{graphicx}
\usepackage{dcolumn}
\usepackage{multirow}
\usepackage{bm}
\usepackage[hypertex,colorlinks=true,linkcolor=black,citecolor=blue,filecolor=blue,urlcolor=blue,setpagesize=false,nesting=true]{hyperref}

\usepackage[utf8]{inputenc}
\usepackage[T1]{fontenc}
\usepackage{mathptmx}

\newcommand{\msr}{$\mu$SR}%

\begin{document}
\title{Ambipolar Property of Isolated Hydrogen in Oxide Materials Revealed by Muon}

\author{M. Hiraishi}
\affiliation{\IMSS}
\author{H. Okabe}
\affiliation{\IMSS}
\author{A. Koda}
\affiliation{\IMSS}\affiliation{\Sokendai}
\author{R. Kadono}\email{ryosuke.kadono@kek.jp}
\affiliation{\IMSS}\affiliation{\Sokendai}\affiliation{\MCES}
\author{H.~Hosono}
\affiliation{\MCES}
\date{\today}

\begin{abstract}
The study on the electronic state of muon as pseudo-hydrogen (represented by the elemental symbol Mu) by muon spin rotation has long been appreciated as one of the few methods to experimentally access the electronic state of dilute hydrogen (H) in semiconductors and dielectrics. Meanwhile, theoretical predictions on the electronic state of H in these materials by first-principles calculations using density functional theory (DFT) do not always agree with the observed states of Mu.  In order to address this long-standing issue, we have re-examined the vast results of previous Mu studies in insulating/semiconducting oxides with special attention to the non-equilibrium character and the ambipolarity of Mu. As a result, we established a semi-quantitative model that enables systematic understanding of the electronic states of Mu in most oxides.
First of all, Mu often occurs simultaneously in a neutral (Mu$^0$) and a diamagnetic state (Mu$^+$ or Mu$^-$) in wide-gap oxides. This is not explained by DFT calculations, as they predict that H is stable only in a diamagnetic state with the polarity determined by the equilibrium charge-transition level ($E^{+/-}$). Our model considers that $\mu^+$ interacts with self-induced {\sl excitons} upon implantation to form relaxed-excited states corresponding to a donor-like (Mu$_D$) and/or an acceptor-like (Mu$_A$) states. Moreover, these states are presumed to accompany the electronic level  ($E^{+/0}$ or $E^{-/0}$) predicted by the DFT calculations for H.  By considering that the stability of these two states including their valence is determined by i) the relative position of $E^{\pm/0}$ in the energy band structure of the host and ii) a potential barrier associated with the transition between Mu$_D$ and Mu$_A$, we find that the known experimental results can be explained systematically in accordance with $E^{\pm/0}$. The model also reveals some common properties of Mu-related defects which were previously regarded as individual anomalies. One is the polaron-like nature of the electronic states associated with shallow donor Mu complexes, for which we argue a major shift of the viewpoint to the Mu$^+_D$-bound excitons. Another is the fast diffusion of Mu$^0_A$, which is understood by the isolated feature of the acceptor-like electronic states.  The possible impacts of these findings to a wide range of insulating compounds is discussed by drawing on examples such as GaN, FeS$_2$, and NaAlH$_4$ as key materials for the green technologies.
\end{abstract}

\maketitle

\section{Introduction}
Hydrogen (H) is conventionally categorized as one of the group 1 elements on the periodic table. This is because H, like alkali metals, readily donates electrons in many redox reactions. On the other hand, it is also well known that H takes a relatively stable anion (hydride) state.  In this case, H can be regarded as an element in the same family as halogens (group 17). Since it became known in the 1980s that H can interact with both $n$-type and $p$-type impurities in silicon and greatly affect the electrical conductivity \cite{Pankove:91,Peaton:92}, H has been attracting considerable attention as a special impurity that exhibits ambipolarity in the field of semiconductors.

It is known from previous studies that most of the incorporated H forms complexes with other impurities and defect centers such as atomic vacancies, thereby causing passivation (loss of electrical activity). Unlike the transfer of electrons between impurity levels (carrier compensation) that occurs when both $n$-type and $p$-type defect centers coexist, the impurity levels themselves are eliminated from the band gap in the passivation due to solid-state chemical reactions. Such complex defects have already been analyzed by various experimental techniques, and their local structures are being clarified.

Meanwhile, another important issue is the electrical activity of H itself as a defect center. It is expected that a few ppm of H (equivalent to about $\sim$10$^{15}$--10$^{16}$/cm$^3$) can be easily incorporated into the material unintentionally during the manufacturing process.  This is comparable to the carrier concentration caused by deliberate impurity addition, thus having a significant impact on conductivity by itself. Including this point, understanding the local electronic state of isolated H is of fundamental importance in elucidating the entire mechanism of contribution for H to electrical activity in semiconductors at the atomic level. However, the amount of isolated H in bulk solids is relatively small, and spectroscopic techniques to investigate its local electronic state are limited.

The muon spin rotation ($\mu$SR) is one of such techniques to obtain the relevant information by implanting a positively charged muon ($\mu^+$, hereafter simply called muon) into the target material and investigating its electronic state as pseudo-H. The muon is an unstable subatomic particle, and it is available as a particle beam in dedicated accelerator facilities.  It behaves as a light radioactive isotope of the proton (with about 1/9 of the proton mass) in terms of chemical properties upon incorporation into matter. This is because the muon mass is two orders of magnitude larger (about 206 times) than the electron mass, and thus making the adiabatic approximation sufficient for understanding muon-electron interaction.  As a matter of fact, the difference in the Bohr radius between a muon binding a single electron, called muonium, and the corresponding neutral H atom is only 0.43\%, so that they can be regarded as having practically the same electronic structure. 

On the other hand, the light mass of muon compared with H assumes relatively large isotope effects on the dynamical properties such as diffusion in solids. For example, the zero-point energy $E_0$ is proportional to the square root of the particle mass in a harmonic potential. Since muon/muonium has nearly 3 times greater $E_0$ than that of H, the activation energy for the former in the over-barrier hopping motion is reduced by $\sim2E_0$. The large zero-point motion also leads to a greater tunneling probability to the neighboring sites, thus enhancing tunneling-mediated diffusion (quantum diffusion). While the present work is mostly concerned with the quasistatic local electronic properties, we will refer to the quantum diffusion of muonium at low temperatures in Sect.~\ref{Mudiff} as an example for the quantum diffusion.

In the discussion on muon as pseudo-H, it is convenient to have the elemental name. Hereafter, the symbol Mu will be used for this purpose (corresponding to H for hydrogen), and the valence states of Mu will be denoted as Mu$^+$, Mu$^0$, and Mu$^-$.

Since the dawn of $\mu$SR research in 1970s, various electronic states and dynamics of Mu have been experimentally revealed in a wide variety of materials including oxides. In addition, the recent progress of first-principles calculations using density functional theory (DFT) in accordance with the advent of computational environment has made it possible to discuss the local electronic structures of Mu/H in individual materials in great detail. Meanwhile, the construction of a physical model that would allow us to understand the Mu states in a cross-material way is still in its infancy. Related with this,  relatively little attention has been payed to the ambipolarity of Mu/H in itself, and the mainstream of research to date has been concerned with the donor-like behavior of Mu/H as a member of group 1 elements. 

In this paper, we will show that the ambipolar property of Mu/H, including its acceptor-like behavior, is the key to the coherent understanding of the local electronic state of Mu. We also argue that the fact that we experimentally observe such ambipolar Mu states is inextricably linked to the another fact that the initial state of Mu is in a temporary non-thermal equilibrium state. The primary goal of this paper is to construct a semi-quantitative model for a unified understanding of the electronic structure of Mu by properly taking into account these two factors. 

To this end, we have compared the experimentally observed electronic states of Mu in various oxides with those predicted by DFT calculations for H in the respective oxides published to date; it is not the scope of this paper to examine the precision of previous calculations by performing new ones by ourselves.   In fact, the results of these existing calculations constitute the basis for our model, and it could be argued that the success of the model serves as evidence that they are sufficiently reliable for our goal.  While we provide a brief summary on the general aspects of the DFT calculations in Sect.~\ref{DFT}, readers are encouraged to refer to the individual  references cited in Table~\ref{Boffset} in Sect.~\ref{muh} for their details.

\section{Notable features of ${\bm \mu}$SR as a method for studying pseudo-hydrogen}\label{nfmusr}

In the actual $\mu$SR experiment, nearly 100\% spin-polarized $\mu^+$ is implanted into a sample, and the time-dependent spatial asymmetry ($\simeq$20 \%) of positrons emitted with high probability in the direction of spin polarization upon beta decay is observed. When $\mu^+$ is implanted into a solid material, it decelerates in a short time (generally less than 1 ns) and comes to rest at an interstitial position.  From that moment until the beta decay occurs (with the mean lifetime $\tau_\mu=2.198$ $\mu$s), $\mu^+$ behaves as Mu to take a variety of valence states according to the local environment. The muon spin exhibits precession at a frequency proportional to the hyperfine interaction between Mu and the surrounding electrons and/or nuclear spins. The hyperfine interaction is described by the Hamiltonian
\begin{equation}
{\cal H}/\hbar = \gamma_\mu{\bm H}({\bm r})\cdot{\bm S}_\mu = \frac{1}{2}[2\pi{\bm A}({\bm r})]\cdot{\bm S}_\mu,
\end{equation}
where $\gamma_\mu$ is the muon gyromagnetic ratio ($=2\pi \times 135.53$ MHz/T), ${\bm S}_\mu$ is the muon spin operator, and ${\bm H}({\bm r})$ is the effective hyperfine field [${\bm A}({\bm r})$ being the hyperfine parameter] at the Mu position ${\bm r}$ in the crystalline lattice \cite{MSR}. Because of the large difference in ${\bm H}({\bm r})$, the paramagnetic state (Mu$^0$) can be readily discerned from the diamagnetic state (Mu$^+$ or Mu$^-$) by the $\mu$SR frequency spectrum (see Appendices A and B for more details). 
In contrast, the distinction between Mu$^+$ and Mu$^-$ needs high-precision chemical shift measurements ($\sim$10$^1$ ppm) \cite{Hiraishi:16}. The time evolution of the muon polarization is observed as a statistical average of the signals from a large number of Mu over a time period of about $10\tau_\mu$ ($\sim$20 $\mu$s), with $t=0$ defined by the time of $\mu^+$ arrival. 

Here, we would like to mention some practically important features for Mu in mimicking H. First, the penetrating power of the muon beam is sufficiently large so that it should not be affected by the surface condition of the sample (bulk-sensitive). The effective concentration of the implanted muons is extremely dilute; even in a pulsed beam experiment where a large number of muons are injected at once, the number of muons present in the sample at the same time is at most $\sim$10$^4$/cm$^2$ per cross section.  The stopping range for muons with a typical incident energy of $T_\mu\simeq4$ MeV is about 0.1--1 mm from the sample surface, so their volume concentration is less than $\sim$10$^5$ muons per cm$^3$. Moreover, it does not accumulate in the sample, as it disappears in a short time ($\sim\tau_\mu$). Therefore, muons provide a unique opportunity to observe the electronic state of pseudo-H in the true dilute limit. 

Meanwhile, as will be explained in detail in Sect.~\ref{must}, the initial state of the implanted Mu is mostly in a relaxed-excited (metastable) state caused by the interaction with the electron-hole pairs (or excitons) generated by the transfer of $T_\mu$ along the muon track to the host lattice. This is also implied by experiments with recently available low-energy muon beam (LEM, $T_\mu\simeq 1$--$30$ keV at Paul Scherrer Institute, Switzerland), in which muons are implanted into a region of $10^1$--$10^2$ nanometers from the sample surface. Although the Mu density in this case is still in the dilute limit, it has been demonstrated that the fractional yields of Mu in different valence states are strongly dependent on $T_\mu$ \cite{Prokscha:07}, indicating the definitive influence of Mu-exciton interaction in determining the final Mu states. 

That the electronic state of Mu does not necessarily correspond to the thermal equilibrium state of H may seem to imply its limitations as a source of information for H. However, as discussed below, it is this non-equilibrium nature that allows us to use Mu to experimentally evaluate the ambipolarity of H. In addition, it should be noted that many electronic materials, including oxides, are used in devices under various electronic excitations such as electric fields and optical irradiation. In this regard, the information obtained from Mu will provide microscopic clues for clarifying the effect of H in those materials on their performance under such electronic excitations (see Sect.~\ref{pol}, for example). Thus, Mu serves as a complementary tool to H itself to reveal the whole picture on the behavior of H in matter.

\section{Mu study--An Approach from Non-thermal Equilibrium States}\label{must}
\subsection{DFT calculations for H defect centers}\label{DFT}
\begin{figure*}[t]
	\centering
	\includegraphics[width=0.97\linewidth,clip]{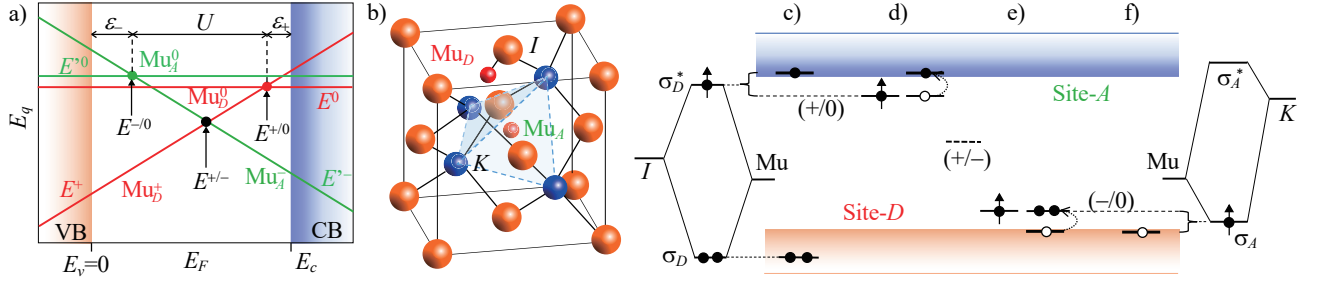}
	\caption{a) Schematic illustrations for the formation energy ($E^q$) of Mu$^q$ ($q=0,\pm$) vs Fermi energy ($E_F$), b) the local structure of a donor-like center (Mu$_D$ at Site-$D$, bonded to an $I$ anion) and an acceptor-like center (Mu$_A$ at Site-$A$, bonded to $K$ cations) in binary compounds, c)--f) the corresponding band diagrams derived from molecular orbital models (where arrows indicate spin-degrees of freedom for localized electrons);  ($+/-$) refers to the equilibrium charge-transition level, ($+/0$) to the donor level (c,d), and ($-/0$) to the acceptor level (e,f).  
}
	\label{mue}
\end{figure*}

In general, the formation energy $E^q$ of H defect centers as a function of the Fermi level $E_F$ is estimated by DFT calculations using the following equation,
\begin{equation}
E^q(E_F)=E_{t}[{\rm H}^q]-E_{t}[\mathchar`-]+qE_F - n_{\rm H}\mu_{\rm H}\label{eqf}
\end{equation}
where $E_{t}[{\rm H}^q]$ and $E_{t}[\mathchar`-]$ denote the total energy of a supercell involving H$^q$ and
a perfect cell, respectively, calculated for charge $q$ ($=\pm,0$), $n_{\rm H}$ is the number of H atoms, and $\mu_{\rm H}$ is
the reference chemical potential for H \cite{Walle:04}. Provided that 
$$E_{t}[{\rm Mu}^q]=E_{t}[{\rm H}^q],$$ 
which is valid within the adiabatic approximation, Eq.~(\ref{eqf}) gives the formation energy for Mu$^q$ as schematically shown in Fig.~\ref{mue}a. 
Although Mu/H can play the role of either a cation or an anion with respect to the host, their local structures can be different with each other (Fig.~\ref{mue}b). Therefore, we refer to them as Site-$D$ (donor-like, associated with anions) and Site-$A$ (acceptor-like, associated with cations). In addition, among the three charge-transition energies,  the equilibrium charge-transition level ($E^{+/-}$) is lower than the acceptor/donor level ($E^{\pm/0}$) in most cases (see Fig.~\ref{mue}a). This behavior is characteristic for systems with strong electron-phonon coupling \cite{Anderson:75},  indicating that the effective onsite Coulomb repulsion energy ($U$) is negative. The negative $U$ character combined with the ambipolarity leads to a tendency of charge disproportionation for H (i.e., preferring H$^\pm$ to H$^0$ states) \cite{Neugebauer:95,Yokozawa:97}. The electronic state of H in the thermal equilibrium is then determined by the relationship among $E^+(E_F)$, $E^0(E_F)$, and $E^-(E_F)$. More specifically, only H$^+_D$ ($E_F<E^{+/-}$) or H$^-_A$ ($E_F>E^{+/-}$) will be realized, and thus $E^{+/-}$ will be the effective impurity level.

In order to predict the electrical activity of H defects by DFT calculations, it is necessary to be able to predict  the band structure of the defect-free host with the accuracy comparable with that for $E^q(E_F)$. However, early DFT calculations exhibited common tendencies that i) they significantly underestimate the band gap, and that ii) the charged defect levels are sensitive to corrections for the finite supercell size, and various correction methods have been developed to mitigate these problems \cite{Lany:08}. Most of the DFT calculations we referred to for oxides were done in the last decade, and they are expected to reflect the results of such developments. Meanwhile, it is also expected that the results of these calculations may be skewed to some extent by differences in the actual prescriptions adopted for the correction. 

In this regard, it is noteworthy that, among the 18 oxides in Table \ref{Boffset} on which our model is based, 9 of them rely on the single set of calculations performed by Li and Robertson \cite{Li:14}.  This may be helpful in assessing the systematic reliability of the DFT calculations.  According to Ref.~\cite{Li:14}, their calculations were carried out using the plane wave pseudopotential code CASTEP \cite{Segall:02}. More specifically, norm-conserving pseudopotentials were used to represent the atomic potentials. The cutoff energy for plane waves was 800 eV.  For the hybrid functional, the one introduced by Heyd-Scuseria-Ernzerhof (HSE06) was employed \cite{Heyd:03,Heyd:06}. To correct the band gap errors of pure Generalized Gradient Approximation (GGA), a fraction $\alpha$ of the short-range separated part of the Hartree-Fock (HF) exchange was combined with the GGA exchange-correlation integral, where $\alpha$ was varied to fit the band gap for systems with a greater gap. The screening length was set to $\mu=0.106$ bohrs$^{-1}$ \cite{Heyd:06}. The calculated band gap energies are quoted in Table \ref{Boffset} for comparison with experimental values.

For the defect calculations in Ref.\cite{Li:14}, the lattice parameters for crystalline oxides were set to experimental values, and only the internal atomic coordinates for the interstitial H were relaxed. The electronic states turned out to be localized enough to save supercell size; they needed supercells with 33-49 atoms each for diverse oxides. The H atom was positioned in an arbitrary location near the open interstitial site's center.  The cutoff energy was 800 eV, and the $k$ point mesh was $2\times2\times2$. The defect formation energies were calculated by Eq.~(\ref{eqf}), where the reference chemical potential was defined by following the method described in Ref.~ \cite{Lany:08}.

\subsection{Mu as relaxed-excited states and acceptor/donor levels}
The attempt to interpret the electronic states of Mu in terms of $E^{+/-}$ fails to explain the existence of the paramagnetic Mu$^0$ state reported in many wide-gap oxides (see Table \ref{Boffset}). This necessitates the introduction of the hypothesis that the initial Mu state immediately after $\mu^+$ implantation to rest corresponds to a relaxed-excited state upon rapid quenching from infinite temperature [i.e., $\beta\equiv1/k_BT\rightarrow0$ in the partition function $Z(\beta)$; see Appnd. C]. 
The relative yields of the final states are determined by the density of states for the available sites and valence states around a common formation energy irrespective of $E_F$.

Here, an intriguing fact to remember for considering the origin of Mu$^0$ is that interstitial paramagnetic H centers (H$^0_i$) are produced by irradiation of H-containing ionic crystals with ultraviolet (uv) light at low temperatures. For example, it is known that  H$^0_i$ (known as $U_2$-centers) is produced in alkali halides containing OH$^-$ defects in the photodissociation reaction,
\begin{equation}
[{\rm OH}^-]+h\nu\rightarrow [{\rm O}^{-}](=[h^+])+{\rm H}^0_i,\label{H0i}
\end{equation}
where [\ ] refers to the anion substitutional site, $h\nu$ to the photons, and $h^+$ to the hole \cite{Kerkhoff:63,Spaeth:70,Morato:80}. This is considered to be a process similar to that of self-trapped-exciton (STE) formation by electronic excitation of the halogen sublattice, $2X^- + h\nu\rightarrow [X^-_2](=[h^+])+ e^-_i$, with $X$ denoting the halogen atoms. (It is known that holes comprise $X^-_2$ dimers in alkali halides \cite{Song:96}.)  Since excited electrons are not self-trapped by themselves \cite{Itoh:97}, the STE formation is understood as the capture (localization) of excited electrons by the Coulomb interaction with self-trapped holes. Thus, it is interpreted that [O$^-$] corresponds to the self-trapped $h^+$, and  H$^0_i$ to $e^-$ captured by H$_i^+$ (in place of $h^+$), respectively.  The H$^0_i$ state in alkali halides is presumed to be stabilized by the antibonding character with halogen atoms \cite{Cho:66} and the bonding character with alkali metals: it must be noted that the excited electron is a dangling bond for the cation.   It is well known that the atomic Mu$^0$ state observed in alkali halides can be regarded as the counterparts of the H$^0_i$ center \cite{Baumeler:86}, where the electronic excitation is induced by the kinetic energy of incident muon; it is estimated that $\sim$10$^3$ $e^-$-$h^+$ pairs (excitons) are produced from 4 MeV muons \cite{Itoh:97}.
Regarding oxides, a process similar to Eq.(\ref{H0i}) has been reported in OH-containing $\alpha$-SiO$_2$ (silica) upon the exposure to ionizing radiations at low temperatures \cite{Tsai:89,Verdi:93,Ichikawa:93,Shkrob:96,Shkrob:97,Shkrob:99}.  Here, it is reasonable to assume that the excited electrons on Si $3p$ orbitals \cite{Ramo:12,El-Sayed:15} are eventually captured by H$^+_i$ (that mimics the role of $h^+$ localized on O).

These observations strongly suggest that H$^0_i$ centers (and corresponding Mu$_A^0$ states) exist as the relaxed-excited state, accommodating the electron in the acceptor level.  In other words, the H$_i^+$ (Mu$_A^+$) state created immediately after the electronic excitation serves as a center of complex formation analogous to the ``acceptor-bound exciton'' \cite{FunSem}. Evidence for the interaction between Mu$^+$ and excitons (not just electrons) is found, for example, in the blueshift ($\sim$0.5 eV) of the luminescence from muon-induced STE's in KBr \cite{Kadono:91,Kadono:92}. This blueshift can be now attributed to the formation of the ionized Mu$^0$-bound exciton, where the luminescence occurs between $h^+$ and $e^-$ bound to Mu$^0$ upon the annihilation of $\mu^+$ by the beta decay, $[X_2^-]\cdot{\rm Mu}^0\rightarrow [X_2^-]\cdot e^-_i\rightarrow 2X^-+h\nu$, where the lattice relaxation for $e^-$ is presumed to be smaller than that for the native STE. 

Considering that Mu in oxides acts as a trapping center for the self-induced free excitons, the initial electronic state of the ambipolar Mu is not limited to the acceptor-like state.  For example, let us examine mono-oxides, $K$O (with $K$ denoting the divalent cations). The free exciton electrons and holes, conveniently expressed as $e^{-*}$, $h^{+*}$, interact with Mu to form the states respectively corresponding to the Mu$_A$ and Mu$_D$ states, i.e.,
\begin{eqnarray}
{\rm Mu}^+ +e^{-*} &\rightarrow& [K^{2+}]\cdot{\rm Mu}^0 = {\rm Mu}^0_A,\\
{\rm Mu}_A^0 +h^{+*} &\rightarrow& [{\rm O^{2-}}]\cdot{\rm Mu}^+ = {\rm Mu}^+_D.
\end{eqnarray}
It is known that the yield of Mu$^0_A$ is actually bottlenecked by the electron supply from ionization trails \cite{Storchak:97,Eshchenko:99,Eshchenko:03,Prokscha:07}.  While both electrons and holes are not self-trapped in many oxides including Al$_2$O$_3$, MgO, ZnO, and crystalline SiO$_2$, holes are self-trapped in $\alpha$-SiO$_2$ and in alkali halides \cite{Itoh:97}. In the latter case, the yield of Mu$^+_D$ may depend on the mobility of Mu$^0_A$  (see Sect. \ref{Mudiff}). 

Mu$^+_D$ is likely to have a good chance of capturing another electron to become Mu$^0_D$  (which may be equivalent with ``donor-bound excitons'').   In fact, there are known examples of donor-like H/Mu defect centers in ionic compounds including oxides. One such classical example in the alkali halides is the hydrogen bifluoride complex, F$^-$-H$^+$-F$^-$. This is also known as a prototype of strong hydrogen bonding observed upon X-ray irradiation \cite{Jeffery:97}.  The analogous F$^-$-Mu$^+$-F$^-$ complexes have been reported in various alkali fluorides and alkaline earth fluorides. They are readily identified by the characteristic $\mu$SR signal due to a well-defined magnetic dipolar field exerted from the two $^{19}$F nuclei  (spin $I=1/2$) \cite{Brewer:86}. 
The donor-like character is evident in the presumed formation process,
Mu$^0$ + $h^{+*}$ (= F$_2^-$) $\rightarrow$ F$^-$Mu$^+$F$^-$.
Note that these donor-like states often coexist with Mu$_A^0$ \cite{Baumeler:86}. 
The occurrence of two different paramagnetic centers corresponding to Mu$_D^0$ and Mu$_A^0$ are well established in elemental (group 4) and group 13-15 compound semiconductors \cite{Patterson:88,Pankove:91}.

This renewed ``radiolysis model'' with emphasis on the Mu-{\sl exciton} interaction not only provides a microscopic model of Mu$^0$ formation but also retain the merit of explaining the finite yield of Mu$^0$ by electron-supply-limited processes. But the model fails to account for the increase in the initial Mu$^0$ yield with increasing temperature in place of the diamagnetic state, e.g., in Lu$_2$O$_3$ \cite{Vilao:18}. This has led to the introduction of the ``thermal spike'' model \cite{Vilao:19}. In this model, the effective local lattice temperature is presumed to be temporally elevated by phonon excitation (within the time scale of sub-picoseconds) around the Mu stopping position. The thermal spike model has a long history of its own since 1960s \cite{Toulemonde:96}, developed to understand the localized damage around ion tracks during irradiation of materials by heavy ions and fission products. The model for the initial Mu states is concerned with the kinetic energy range lower than that leading to atomic displacements by the knock-on processes. 

\begin{figure}[t]
	\centering
	\includegraphics[width=0.88\linewidth,clip]{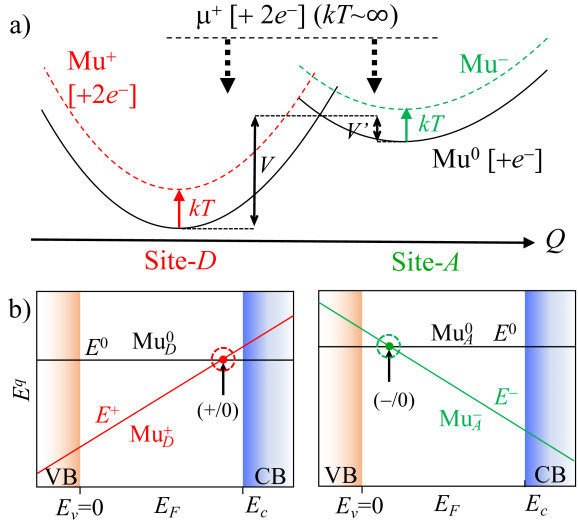}
	\caption{a) Adiabatic potential curves for Mu in binary compounds, where each curve represents Sites-$D$ and -$A$ with different valences. (NB: those for Mu$^\pm$ depends on $E_F$.) If there is a potential barrier between Site-$D$ and -$A$ (e.g., $V$ and $V'$ between Mu$^0$ states) , both are observed simultaneously ($Q$ is the configuration coordinate).  b), c)  The electronic states of Mu relevant to the respective sites are shown by dashed circles in the $E^q$ vs $E_F$ diagram.}
\label{adia}
\end{figure}

Considering the above discussions, we assume that the initial state of Mu is determined roughly through two steps. First, immediately after coming to rest, Mu forms ambipolar relaxed-excited states represented by an adiabatic potential  shown in Fig.~\ref{adia}a \cite{Itoh:97}. The critical hypothesis here is that these states correspond to Mu$^0_A$ and/or Mu$^0_D$ in Fig.~\ref{mue}a originally predicted for H by the {\it ab initio} DFT calculations. The situation can be described as that attained by the temporary shift of $E_F$ from thermal equilibrium to the region $E^\pm(E_F)>E^0(E_F)$. The variation of these states with temperature is then interpreted to reflect the degree of relaxation for $E_F$ from around $E^{\pm/0}$ towards $E^{+/-}$ within the observation time ($<10^{-5}$ s) at each temperature.   For instance, regarding Mu$_D$ in Fig.~\ref{adia}b, the transition from Mu$_D^0$ to Mu$_D^+$ occurs as  the temporal $E_F$ decreases from the middle between $E^{+/0}$ and $E_c$ to the equilibrium level (either $E^{+/-}$ of Mu/H or other impurities, leaving the $E^{+/0}$ level empty) with increasing temperature ($k_BT\gg \epsilon_+$), which is interpreted as the promotion of an electron from the $E^{+/0}$ level to the conduction band (see Fig.~\ref{mue}d). Therefore, if $E^{+/0}$ is located within the band gap, Mu$^0_D$ can be realized as the initial state. Meanwhile, if $E^{+/0}$ is in the conduction band ($E^{+/0}>E_c$) and there is no barrier associated with charge conversion, Mu$^0_D$ will immediately ionize and take the Mu$^+_D$ state (Fig.~\ref{mue}c), meaning that it behaves as an $n$-type impurity regardless of temperature.  The same is true for the $p$-type activity of Site-$A$ (Figs.~\ref{adia}c, \ref{mue}e and \ref{mue}f). 

Moreover, if the Mu$_D$ and Mu$_A$ states are separated by an energy barrier $V$ (and $V'$) on the frame of configuration coordinate (Fig.~\ref{adia}a), then Mu can take two corresponding electronic states as initial states at low temperatures ($V,V'\gg k_BT$). The yield of each state is proportional to the relative density of states which also depends on temperature (see Appnd. C). Recently, an attempt has been made to evaluate this potential from experimentally observed yields of Mu$^0$ and Mu$^+$ in Lu$_2$O$_3$, assuming that the relative yields of these states are determined by the potential similar to that shown in Fig.~\ref{adia}a, within a short time from muon stopping ($\sim$10$^{-12}$ s) to the completion of the lattice relaxation ($\sim$10$^{-10}$ s) \cite{Vilao:17}. When $V\le0$ or $V'\le0$, only one of these will be realized as the initial state, and its ionization is observed with increasing temperature.  Since such initial states cannot be readily realized in an experiment for H under normal conditions, it is a major advantage of Mu study to allow the direct access to donor/acceptor levels ($E^{\pm/0}$). 

\section{Mu in insulating/semiconducting oxides}

\begin{figure*}[t]
	\centering
	\includegraphics[width=0.75\linewidth,clip]{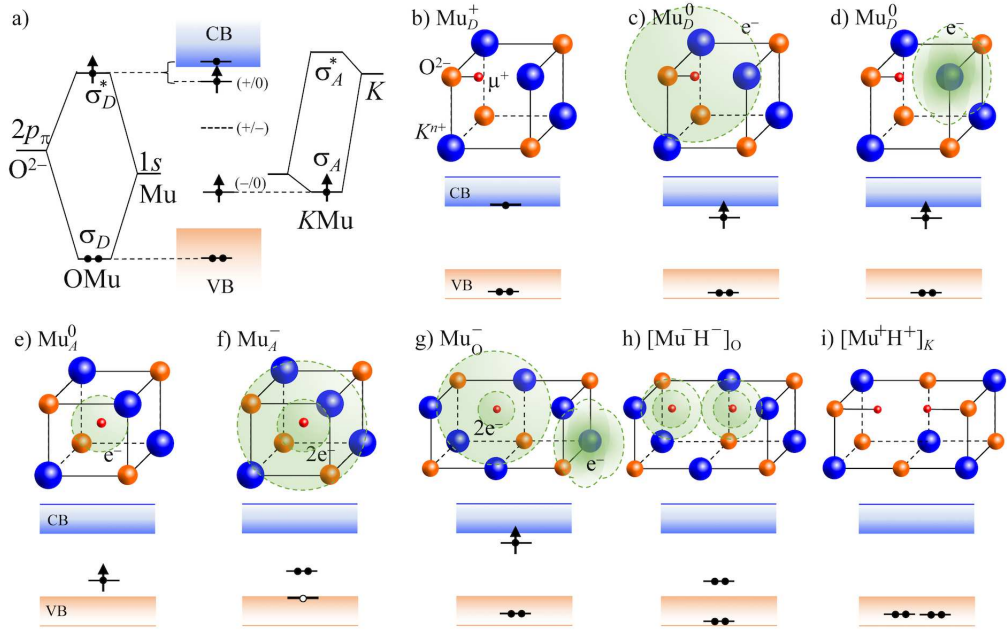}
	\caption{a) Schematic illustration of the relationship between electronic levels and band diagrams in oxides based on molecular orbitals with Mu forming chemical bonds with oxygen (left) or cations (right).  b--i) Typical defect structures involving Mu and the associated electronic states; those in b--d) indicate the states where OMu bonds are formed (Mu$_D$ in Fig.~\ref{mue}b), e--f) indicate the states where Mu forms multiple bonds with cations (Mu$_A$ in Fig.~\ref{mue}b), g--h) indicate complex states involving oxygen defects, and i) indicates that associated with divalent cation vacancies.  d) and g) are polaron states, where electrons are localized in cationic electron orbitals (e.g., $d$ orbitals). b--d) and g) can have $n$-type activity, whereas e--f) can be $p$-type active.
	}
	\label{muc}
\end{figure*}

\subsection{Typical electronic structure of Mu-containing defects revisited from the viewpoint of ambipolarity}\label{Mustat}
In oxides with strong ionic bonding, the bottom of the conduction band consists mainly of the cation $s$ band and the top of the valence band consists of the $2p$ band of oxygen. Therefore, Site-$D$ and -$A$ in Fig.~\ref{mue} can be qualitatively understood as the state of Mu governed by the interaction with the ligand oxygen (O$^{2-}$) and cation ($K^{n+}$), respectively. The schematic local structure of Mu specifically assumed here is shown in Fig.~\ref{muc}. The typical state at Site-$D$ is that associated with the formation of OH bonds with oxygen by interstitial Mu (Fig.~\ref{muc}b--d). If we consider the molecular orbitals in Fig.~\ref{muc}a, the bonding orbital between O $2p$ and Mu $1s$ ($\sigma_D$) is filled with two covalent electrons and sinks to a deep position in the valence band, while the antibonding orbital ($\sigma^*_D$) is pushed up to the conduction band.

The state of the remaining one electron is determined by the hybridization of the $\sigma^*_D$ orbital and the conduction band, and if the hybridization with $s$-$p$-like orbitals is strong, the electron enters the bottom of the conduction band ($E_c$), which contributes to conductivity regardless of temperature; O$^{2-} + {\rm Mu}^0$ $\rightarrow{\rm OMu}^- (= {\rm Mu}^+_D) + e^-$ (Fig.~\ref{muc}b). Meanwhile, when the band gap ($E_g\equiv E_c-E_v$) is large and hybridization is weak, Mu strengthens the character of the isolated center and $E^{+/0}$ is formed within the gap (Fig.~\ref{muc}c, d). Even in this case, when $E^{+/0}$ is close to the bottom of the conduction band ($E_c-E^{+/0}\lesssim k_BT$), the bound electron of Mu$^0_D$ is thermally excited to the conduction band at a finite temperature, and the valence state change is observed at elevated temperatures.

On the other hand, in the case of Site-$A$ in Fig.~\ref{muc}a (right), Mu is surrounded by cations (see also Fig.~\ref{muc}e). This state is expected to be stabilized by the formation of multiple bonding with cations \cite{Li:14} (and/or by the antibonding character of hybridization with oxygen 2$p$ band, as discussed for H$^0_i$ in alkali halides), accompanying the acceptor level. The electron can enter the $K$Mu molecular bond orbitals, and serves as an acceptor; when the associated bond level ($\sigma_A$) is close to the valence band ($E^{-/0}-E_v\lesssim k_BT$), Mu is promoted to a hydride-like state by accommodating the second electron and supplying a hole to the valence band top (Fig.~\ref{muc}f), i.e., $K$Mu$^0\rightarrow K$Mu$^-$ (= Mu$^-_A$) $+h^+$. 
When the corresponding electron level is situated near the center of the gap, Mu is observed as an atomic Mu$^0_A$. As discussed below, the Mu$^0$ states observed in wide-gap oxides exhibiting large hyperfine parameters are interpreted as this state, where the multipolar interaction with cations is weakest.
The equilibrium charge-transition level $E^{+/-}$ is located in the middle of $E^{+/0}$ and $E^{-/0}$.

The study of Mu in oxides began on insulators with large $E_g$, such as SiO$_2$ \cite{Myasishcheva:68,Spencer:84}, Al$_2$O$_3$ \cite{Minaichev:70}, or MgO \cite{Spencer:84,Kiefl:86}, where atomic Mu$^0$ states were observed with hyperfine interactions as large as that in vacuum [$|{\bm A}|=A_{\rm vac}= 4463.30$  MHz, see Eq.~(\ref{Amu}) in Appnd.~B]. In subsequent studies, atomic Mu$^0$ has been observed almost without exception in insulators with $E_g$ generally above 6--7 eV \cite{Cox:06b,Marinopoulos:17,Vieira:14,Vieira:16,Silva:12,Silva:16}. However, there are two problems in understanding these experimental results. One is that, as we have already discussed, H$^+$ or H$^-$ is always more stable than H$^0$ irrespective of $E_F$ due to its ambipolarity, which is inconsistent with the fact that Mu$^0$ is observed (given that Mu were also in thermal equilibrium). Another point is that in many of these materials, the diamagnetic Mu (Mu$^+$ or Mu$^-$) is observed to coexist with Mu$^0$ (e.g., in SiO$_2$, the yields of Mu$^0$ and Mu$^\pm$ are $\sim$65\% and $\sim$35\%, respectively \cite{Spencer:84}), but the origin of these diamagnetic components remains to be identified.

These two problems can be resolved by considering that the observed electronic states of Mu correspond to the non-equilibrium states near $E^{\pm/0}$, and that Mu can simultaneously take on donor- and acceptor-like states shown in Fig.~\ref{muc}. In the case of SiO$_2$, the position of $E^{\pm/0}$ inferred from DFT calculations for H suggests that the diamagnetic state and Mu$^0$ state correspond to Mu$^+_D$ (Figs.~\ref{muc}b) and Mu$^0_A$ (Figs.~\ref{muc}e), respectively (see Section \ref{muh} for more details).

In contrast to the above, the first example of Mu$^0$ with a ``shallow donor level'' ($0<\epsilon_+\ll E_g$) was found relatively recently in ZnO \cite{Cox:01,Shimomura:02}. This discovery was made in response to the prediction of an {\it ab initio} DFT calculation where both $E^{+/-}$ and $E^{\pm/0}$ lied $\sim$0.4 eV below $E_c$  \cite{Walle:00},  leading to the widespread search for shallow donor H/Mu in oxides. According to the earlier report of the $\mu$SR study on powder ZnO samples, a single Mu$^0$ state was observed with the hyperfine parameter described by $A(\theta,\phi)=A^*+D\cos^2\theta$ with $A^*=0 .50(2)$ MHz and $D=0.26(2)$ MHz, respectively \cite{Cox:01}. Here, $\theta$ ($\phi$) is the polar  (azimuthal) angle with respect to the symmetry axis of $A$ [see Eq.~(\ref{Amu2}) in Appnd.~B for more details].
  
Apart from the fact that the values of $A^*$ and $D$ are comparable and thus the hyperfine interaction is clearly anisotropic, the value of $A^*$ is orders of magnitude smaller than that of atomic Mu$^0$ in vacuum ($A^*/A_{\rm vac}\simeq10^{-4}$), leading to the consensus that the electronic state is qualitatively understood by the effective mass model with a large Bohr radius [corresponding to Fig.~\ref{muc}c, see Eq.~(\ref{efm}) in Appnd.~B]. Here, in order to quantitatively evaluate the origin of the hyperfine interaction, we consider the Fermi contact term $A_c$ and the dipole field $A_d$ from the localized moment on the symmetry axis, so that the hyperfine parameters are expressed in the following form, 
\begin{equation}
A(\theta,\phi)=A_c+\frac{A_d}{2}(3\cos^2\theta-1), \label{Amu3}
\end{equation}
where the second term corresponds to the case where the principal axis of the tensor $\hat{A}_{\rm d}$ representing the dipole field from the electron is taken in the $z$ direction [see Eq.~(\ref{HMu}) in Appnd.~B]. Then, from the relations $A^*=A_c-A_d/2$ and $D=3A_d/2$, we obtain $A_{c}=0.579(9)$ MHz and $A_{d}=0.177(5)$ MHz.
These values suggest that the electrons associated with Mu$^0$ are rather close to the intermediate situation between Figs.~\ref{muc}c and d. Here, the electron responsible for $A_d$ is assumed to be  in the 4$s$ orbital of Zn with considerable degree of delocalization. The fact that $A_d$ takes a value comparable to $A_c$ suggests the formation of an off-center polaron state in which the centers of positive and negative charges do not match.

Subsequent measurements on single crystals revealed that there were two different Mu$^0$ states, where the angular dependence of the hyperfine interaction was isotropic with respect to the rotation of the crystal (wurtzite type) around the $\langle0001\rangle$ axis. This suggests that the local structure of these two Mu$^0$ states respectively correspond to the bond-center and antibonding positions along the Zn-O bond that is parallel with the $\langle0001\rangle$ axis \cite{Shimomura:02} in line with theoretical predictions \cite{Walle:00}.  Looking back from the viewpoint of ambipolarity, the two observed states may correspond to donor/acceptor-like states, where the bond-center Mu$^0$ is tentatively assigned to Mu$^0_D$, and another situated at the antibonding position surrounded by Zn to Mu$^0_A$. This also rises an additional issue on the origin of the diamagnetic state coexisting with these two Mu$^0$ states at low temperatures, which is discussed later (see Sect.~\ref{pol}).

The Mu$^0$ state accompanying a shallow donor-like state observed in TiO$_2$ (rutile) is a typical example corresponding to the limit of the off-center polaron state ($A_d\gg A_c$, Fig. \ref{muc}d). This is a complex state involving Mu, O, and Ti in which the accompanying electron is loosely localized in the 3$d$-orbitals of neighboring Ti atom(s) \cite{Shimomura:15,Vilao:15}. An earlier ENDOR study reported the H-related paramagnetic center with a similar electronic structure in a reduced sample \cite{Brant:11}. In such an electronic state, the hyperfine interaction is dominated by the magnetic dipole interaction [the second term of Eq.~(\ref{Amu3})], so that the hyperfine parameters may satisfy the relation ${\rm Tr}\hat{A}_{\rm d}=0$. In fact, this relationship is nearly satisfied for TiO$_2$, although there are slight differences among the literature \cite{Shimomura:15,Vilao:15,Brant:11}. Notably, the size of the localized moment estimated from $A_d$ is only about $\sim$0.05$\mu_B$, and it is highly likely that the state is more extended (larger polaron-like, or of greater distance between Mu$^+$ and $e^-$). Moreover, the emergence of the second Mu$^0$ states with a greater $A_d$ below $\sim$5 K coexisting with the diamagnetic state \cite{Vilao:15} recalls the situation in ZnO.  Recent studies on Mu in SrTiO$_3$ have also reported electronic states similar to TiO$_2$ \cite{Ito:19}, where the localized moment at the Ti site is as large as $\sim$0.33$\mu_B$, suggesting a more strongly localized state (small polaron-like) than that in TiO$_2$. 

Interestingly, all of the charge transition levels ($E^{+/-}$, $E^{-/0}$, and  $E^{+/0}$) for H in ZnO, TiO$_2$, and SrTiO$_3$ inferred from previous first-principles DFT calculations lie within the conduction band \cite{Walle:00,Li:14,Iwazaki:10}.  The naive application of our model to those oxides would predict only Mu$^+$, which is in contrast to the experimental observations. These disparities have previously been considered as individual anomalies in view of the relative accuracy of the DFT calculations, but we will show in Sect.~\ref{pol} that considering the polaronic state leads to an alternative model for the origin of these shallow levels.

For the sake of completeness, let us now consider the complex state with atomic vacancies. The requirement for stabilizing hydride states is expected to be readily satisfied in oxygen vacancies (V$_{\rm O}^{2-}$).  When their concentration is sufficiently high, Mu may also have a chance to be trapped there to form various local electronic states. For H, it is theoretically predicted that one or two H$^-$ ions are accommodated in a single oxygen vacancy. While the hydride state for the interstitial Mu/H is attained by capturing an electron to the acceptor levels,  the single Mu$^-$/H$^-$ atom trapped in V$_{\rm O}^{2-}$ can be donor-like due to the occurrence of the excess electron, e.g., Mu$^0$+V$_{\rm O}^{2-}\rightarrow$[Mu$^{-}]_{\rm O}+e^-$, where [\ ]$_{\rm O}$ denotes the oxygen substitutional site (see Fig.~\ref{muc}g).  The donor-like behavior of [H$^{-}]_{\rm O}$ is known to be crucial for the realization of high electron doping by O substitution with H, e.g., in iron-based superconductors \cite{Iimura:12}.

On the other hand, when two H$^-$ ions are included, as in Fig.~\ref{muc}h, passivation by charge compensation occurs. This may be equivalent to the formation of metal hydrides. As already mentioned in Sect.~\ref{nfmusr}, the probability of two Mu atoms coming into close proximity under practical experimental conditions is negligibly small. But in situations where there are many oxygen vacancies containing a single H$^-$, Mu is expected to have a good chance of finding such a site. This allows Mu to form a Mu-H complex defect to passivate the electrical activity (Mu$^0$+[H$^{-}]_{\rm O}+e^-\rightarrow$[Mu$^-$H$^-]_{\rm O}$).
In fact, recent muon experiments performed on amorphous InGaZnO$_{4-\delta}$ doped with large amounts of H (a-IGZO:H) have found evidence for the formation of such a state \cite{Kojima:19}. The $\mu$SR spectra observed in IGZO without H-doping, regardless of whether it is crystalline or amorphous, show depolarization described by a Gaussian Kubo-Toyabe function of the diamagnetic Mu [Eq.~(\ref{gkt}) in Appnd.~A], implying that Mu feels a Gaussian random local fields induced by a number of In/Ga nuclear magnetic moments at nearly equal distances.  In contrast, those in a-IGZO:H exhibit a Lorentzian lineshape with an enhancement in the depolarization rate, strongly suggesting the presence of H in addition to In/Ga nuclei at the nearest neighbor of Mu. The latter is in line with the formation of [Mu$^-$H$^-]_{\rm O}$ complex state in a-IGZO:H, where the associated electronic levels are predicted to lie just above the valence band top to accommodate two electrons (see Fig.~\ref{muc}h) \cite{Kojima:19,Li:18}. 
These levels are the prime suspect of the negative bias illumination stress (NBIS) instability for a-IGZO \cite{Nomura:08,Nomura:11,Jeong:13}, where the instability is essentially a persistent photoconductivity caused by the photo-excitation of these electrons into the conduction band. A similar [Mu$^-$H$^-]_{\rm O}$ complex state has been reported in partially hydrated BaTiO$_{3-x}$H$_x$ \cite{Ito:17,Iwazaki:14}.

More interesting point in IGZO from the viewpoint of ambipolarity is that the role of H varies qualitatively with its concentration. As mentioned above, Mu in crystalline and as-deposited amorphous thin films does not form a Mu$^0$ state regardless of temperature. Based on a comparison of the observed magnitude of the $\mu$SR linewidth $\Delta$ [proportional to the mean square of the random field from the nuclear magnetic moments $\langle|{\bm H}_{\rm d}|^2\rangle$, see Eq.~(\ref{delta_n}) in Appnd.~A] and the $\Delta$ predicted for the stable site by DFT calculations, the Mu site was narrowed down to be located in the bond center of ZnO, taking the Mu$^+$ state \cite{Kojima:19}. This means that the local electronic structure of isolated Mu/H in IGZO is represented by that shown in Fig.~\ref{muc}b, which is evidence that H behaves as an $n$-type impurity regardless of temperature in the limit of dilute concentration.  Thus, the ambipolar property of H allows itself to play the roles of donor- and acceptor-like defects, determined by the local atomic structures that depend on the H concentration. 

Finally, regarding the relationship with the cation vacancy, H/Mu serves to compensate the oxygen dangling bonds.  Mu may be found together with other protons in multivalent cation vacancies; e.g., [Mu$^+$H$^+$]$_K$ (see Fig.~\ref{muc}i). This, for example, corresponds to the state known as the Ruetschi proton in manganese oxides \cite{Ruetschi:84,Ruetschi:88}. A recent muon study on rutile-type MnO$_2$ ($\beta$-MnO$_2$) suggests that about 15\% of the implanted Mu are involved in (or located near) the [4H$^+$]$_{\rm Mn}$ complexes, while the remaining 85\% are in the diamagnetic state (OMu$^-$) at the oxygen channels \cite{Okabe:21}.
In this study, hydrogen-sensitive temperature-desorption measurements revealed that as-prepared MnO$_2$ sample contains a considerable amount of H ($\simeq3\times10^{19}$ cm$^{-3}$), indicating that hydrogen impurity can be the origin of the conductivity of $\beta$-MnO$_2$ which is presumed as an insulator ($E_g\simeq0.04$--0.17 eV)

\subsection{Electronic state of implanted Mu determined by\\ ${\bm E}^{+/0}$ and ${\bm E}^{-/0}$}\label{muh}
\begin{figure}[t]
	\centering
	\includegraphics[width=0.95\linewidth,clip]{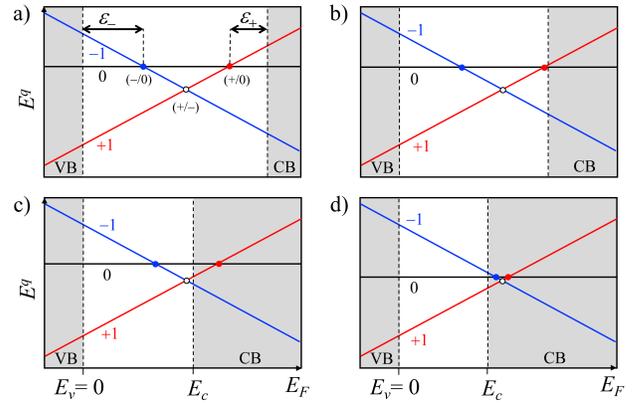}
	\caption{Fermi level dependence of the formation energy $E^q$ of the charged state $q$ ($=0,\pm1$) obtained by DFT calculations for interstitial H in oxides. The symbols $(+/-)$ and $(\pm/0)$ correspond to the mutual intersection points of $E^q$ ($E^{+/-}$ and $E^{\pm/0}$).  The electrical activity of H is determined by the relationship between $E^{+/-}$ and $E_c$ and $E_v$, whereas that of Mu is governed by $E^{\pm/0}$ and is predicted by the sign and value of $\epsilon_\pm$ (see text). }
	\label{Es}
\end{figure}

The results of DFT calculations performed to date on the interstitial H in various oxides can be qualitatively classified into four patterns in terms of the $E_F$ dependence of $E^q(E_F)$, and the relationship between $E^{+/-}$, $E^{\pm/0}$ and the band structure, as indicated in Fig.~\ref{Es}. Provided that $E^{+/-}$ serves as the pinning level for $E_F$ (i.e., a considerable amount of H is present), the electric activity of H is determined by the relationship between $E^{+/-}$ and $E_c$ (measured from $E_v=0$). Fig.~\ref{Es}a-b shows the deep $E^{+/-}$ level, Fig.~\ref{Es}c shows the shallow $E^{+/-}$ level, and Fig.~\ref{Es}d shows that there is no level in the gap and only H$^+$ is stable.  Interestingly, early DFT calculations proposed a model in which $E^{+/-}$ is aligned at a certain energy measured from the vacuum level ($E^{+/-}-E_{\rm vac}\sim-3$ eV) regardless of oxides \cite{Kilic:02}. This model has been also applied to the cases of Mu in the literature \cite{Cox:06b,Cox:06a}, and the existence of shallow donor levels have been predicted in Bi$_2$O$_3$, HgO, Sb$_2$O$_3$, and so on, in which no such state has been actually observed. However, as already mentioned, a coherent interpretation becomes possible for Mu by considering $E_F$ vs $E^{\pm/0}$ instead of $E^{+/-}$ and that Mu is in the relaxed-excited states related with $E^{\pm/0}$.
 
A seminal example for the importance of considering $E^{\pm/0}$ is the early studies of Mu in strongly covalent semiconductors, where both donor- and acceptor-like Mu$^0$ states (respectively corresponding to Mu$^0_D$ and Mu$^0_A$ in Fig.~\ref{mue}b) have been observed to coexist \cite{Patterson:88,Pankove:91}. 
Interpreting the charge state changes of Mu$_D^0$ and Mu$_A^0$ as respectively associated with $E^{+/-0}$ and $E^{-/0}$, and estimating $E^{+/-0}$ by their interpolation using Eq.~(\ref{eqf}), it was found that $E^{+/-}$ was situated near the charge neutral level ($E_{\rm CNL}$) common to materials in question \cite{Lichti:08}.   In this study, the authors focused on whether the position of $E^{+/-}$ is independent of the material (see Sect.~\ref{ImpH} for details), but the results support that the change transition (ionization) of Mu$^0$ depends on the $E^{\pm/0}$ level rather than $E^{+/-}$; note that $E^{\pm/0}$ were also in semi-quantitative agreement with the predictions of DFT calculations \cite{Walle:03}.  Let us now investigate whether a coherent understanding of the Mu valence state in oxides is attained by this presumption.


\begin{table*}[t]
\begin{center}
\begin{tabular}{c|ccc||ccccccl} 
\hline\hline
\multirow{2}{*}{Mater.} & \multicolumn{3}{c||}{Exp. [$E_g$ (eV)]} & \multicolumn{7}{c}{DFT Calc. [$E_g$, $E^{+/-}$, $E^{\pm/0}$, $\epsilon_+$ (eV)] }  \\
		 & $E_g$  & Mu & Refs. & $E_g$ & $E^{-/0}$ & $E^{+/-}$ & $E^{+/0}$ & $\epsilon_+$ & Mu &Refs. \\
\hline
BeO 					& 10.6 &  Mu$^0_X$ &\cite{Cox:06b,Marinopoulos:17} & 10.6 & 5.5 & 6.09 & 6.7 & 3.9 & Mu$^0_{A/D}$ &\cite{Marinopoulos:17}  \\
SiO$_2$				& 9.0 &  Mu$^0_X$, Mu$^{+}$ & \cite{Myasishcheva:68,Spencer:84} & 8.7 & 2.9 & 5.4 & 7.9 & 0.8 & Mu$^0_{A}$, Mu$^0_{D}$$^\bigstar$ & \cite{Li:14} \\
$\alpha$-Al$_2$O$_3$ 	& 8.8 &  Mu$^0_X$, Mu$^{+}$ & \cite{Minaichev:70,Kiefl:84} & 8.5 & 3.0 & 5.4 & 7.7 & 0.8 & Mu$^0_{A}$, Mu$^0_{D}$$^\blacklozenge$ &\cite{Li:14} \\
MgO					& 7.8 &  Mu$^0_X$, Mu$^{+}$ & \cite{Spencer:84,Kiefl:86} & 7.5 & 2.8 & 5.4 & 7.9 & $-0.4$ & Mu$^0_A$, Mu$^{+}_D$ &\cite{Li:14} \\
m-HfO$_2$ 			& 6.0 &  Mu$^0_X$, Mu$^{+}$  & \cite{Vieira:14} & 5.8 & 1.6 & 4.0 & 6.3 & $-0.5$ & Mu$^0_A$, Mu$^{+}_D$ &\cite{Li:14}  \\
q-GeO$_2$ 			& 6.0 &  Mu$^0_X$, Mu$^{+}$  & \cite{Cox:06b,Spencer:84} & 5.6 & 2.0 & 4.8 & 7.2 & $-1.6$ & Mu$^0_A$, Mu$^{+}_D$ &\cite{Li:14}  \\
Lu$_2$O$_3$			& 5.6(1) &  Mu$^0_X$, Mu$^{+}$  & \cite{Silva:16}  & 4.0 & 1.1 & 2.46 & 4.2 & $-0.2$ & Mu$^0_A$, Mu$^{+}_D$ & \cite{Silva:16} \\
ZrO$_2$ 				& 5.5(3) &  Mu$^0_X$, Mu$^{+}$  & \cite{Vieira:16}  & 5.4 & 2.1 & 3.5 & 4.8 & 0.6 & Mu$^0_{A}$, Mu$^0_{D}$$^\bigstar$ &\cite{Marinopoulos:12} \\
Y$_2$O$_3$ 			& 5.5 & Mu$^0_A$, Mu$^{+}$ & \cite{Silva:12} & 5.9 & 2.15 & 3.8 & 5.5 & 0.4 & Mu$^0_{A}$, Mu$^0_{D}$$^\bigstar$ & \cite{Silva:12} \\
La$_2$O$_3$			& 5.4(1) &  Mu$^0_X$, Mu$^0_{S}$ & \cite{Cox:06b}  & 5.2 & 0.3 & 3.0  &  6.2 & $-1$ & Mu$^0_A$, Mu$^{+}_D$\P  &\cite{Li:14} \\
$\beta$-Ga$_2$O$_3$ 			& 5.0 &  Mu$^+$  & \cite{King:10} & 4.8 & 3.2 & 4.9 & 6.4 & $ -1.6$ & Mu$^+_D$ &\cite{Li:14}  \\
c-IGZO			& 3.68 & Mu$^+$ & \cite{Kojima:19} & 3.1 & $>$3.1 & $>$3.1 & 4.8 & $-1.7$ & Mu$^{+}_D$  &\cite{Li:18}  \\
SnO$_2$	 			& 3.6 & Mu$^+$ & \cite{Cox:06b,King:09b} & 3.6 & 4.1 & 4.3 & 4.6 & $-1$ & Mu$^+_D$ &\cite{Li:14} \\
ZnO					& 3.4 & Mu$^0_{S}$, Mu$^+$ & \cite{Cox:01,Shimomura:02} & 3.4 & $\ge$3.4 & $\ge$3.4 & 3.4 & $0$ & Mu$^{+}_D$\P  &\cite{Oba:08}  \\
$\alpha$-TeO$_2$		& 3.4 &  Mu$^0_A$, Mu$^{+}$ & \cite{Vilao:11} & 2.82 & 0.8 & 2.2 & 2.82 & 0 & Mu$^0_A$, Mu$^+_D$ &\cite{Vilao:11}  \\
SrTiO$_3$ 			& 3.2 &  Mu$^0_{S}$, Mu$^+$  & \cite{Salman:14,Ito:19}  & 3.1 & $ >$4 & 3.8 & 3.5 & $-0.4$ & Mu$^{+}_D$\P &\cite{Iwazaki:10,Varley:14} \\
r-TiO$_2$ 			& 3.2 &  Mu$^0_{S}$, Mu$^+$  & \cite{Shimomura:15,Vilao:15} & 3.0 & 2.6 & 3.1 & 3.5 & $-0.5$ & Mu$^{+}_D$\P  &\cite{Li:14}  \\
In$_2$O$_3$			& 2.7(1) & Mu$^{+}$ & \cite{King:09b}  & 2.67 & 1.8 & 3.2 & 3.67 & $-1$ & Mu$^+_D$ &\cite{Limpijumnong:09} \\
\hline
w-GaN					& 3.48 & Mu$^0_S$, Mu$^+$ & \cite{Shimomura:04}  & 3.4 & 1.2 & 2.4 & 3.5 & -0.1 & Mu$^+_D$\P &\cite{Limpijumnong:03} \\
\hline\hline
\end{tabular}
\caption{Comparison between the valence state of Mu in various oxides and DFT calculations for the interstitial H. $E^{\pm/0}$, $E^{+/-}$ are the energy with $E_v$ as the origin ($E^{-/0}=\epsilon_-$). 
Mu$^0_X$ refers to the deep level Mu$^0$ [corresponding to either donor ($X=D$) or acceptor ($X=A$) in DFT calculations], and Mu$^0_{S}$ to the shallow donor level Mu$^0$. For these marked by $\bigstar$, the discrepancy can be attributed to $E^{+/0}$ over indirect gaps around the $\Gamma$ point (see text and Fig.~\ref{BA}) or to the ambiguity in the experimental background mimicking Mu$^+$ signal, while those marked by \P\  exhibits a correlation with the polaronic (off-center) Mu$^0_{S}$. For the case of $\alpha$-Al$_2$O$_3$ (marked by $\blacklozenge$), see text.}\label{Boffset}
\end{center}
\end{table*}

\begin{figure}[t]
	\centering
	\includegraphics[width=\linewidth,clip]{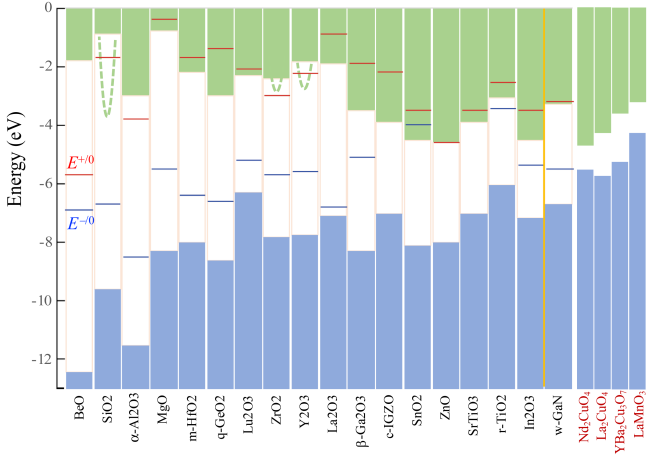}
	\caption{The donor/acceptor levels ($E^{+/0}$/$E^{-/0}$) in Table \ref{Boffset} plotted on the band structure aligned to the vacuum level. The parabolic curves just beneath the conduction band minimum in SiO$_2$, ZrO$_2$, and Y$_2$O$_3$ indicate the presence of dispersive bands with narrower gaps around the $\Gamma$ point. $E^{-/0}$ is unknown for c-IGZO, ZnO, and SrTiO$_3$.  ($\alpha$-TeO$_2$ is not included because the electron affinity is unknown.) The band structure for the typical cuprates/manganites are also shown for comparison (see Sect.~\ref{Impl}).}
	\label{BA}
\end{figure}

In the left column of Table \ref{Boffset}, the experimental results of Mu in several oxides for which DFT calculations have been performed are listed in descending order of the magnitude of $E_g$, and the observed electronic states of Mu are shown. The corresponding energies $E_g$, $E^{\pm/0}$, and $\epsilon_\pm$ (see Fig.~\ref{Es}a for the definition) obtained by DFT calculations are shown in the right column. These compounds comprise a subset of oxides in which the conduction and valence bands consist of empty cation $ns^0$ and O 2$p$ orbitals (represented by Fig.~\ref{muc}a).  Fig.~\ref{BA} shows $E^{\pm/0}$ vs band structure, where all the energy levels are aligned to the vacuum level by considering the electron affinity \cite{Li:14}.   Although it is not explicitly shown in Fig.~\ref{Es}, the local structures of Mu corresponding to $E^q(E_F)$ can be different between $q=+1$ and $-1$ (e.g., Y$_2$O$_3$ \cite{Silva:12}). Therefore we refer to Mu$_D^+$ and Mu$_A^-$ according to  Figs.~\ref{muc}b--f, respectively. The electronic state of Mu observed in the vicinity of $E^{+/0}$ is then assigned to Mu$^0_D$/Mu$^+_D$, and that in the vicinity of $E^{-/0}$ is to Mu$^0_A$/Mu$^-_A$. Based on the comparison between the $E^{\pm/0}$ levels and band structures (which are visualized in Fig.~\ref{BA}), the second column from the right predicts the state of Mu observed as the initial state at low temperatures.  Provided that the transition barrier between Mu$_D$ and Mu$_A$ ($V$, $V'$ in Fig.~\ref{adia}) is sufficiently large, both Mu$^0_{D}$ and Mu$^0_{A}$ may be observed for the case of $\epsilon_+>0$ and $\epsilon_->0$ (Fig.~\ref{Es}a--b), Mu$^0_A$ and Mu$^+_D$ for $\epsilon_+<0$ and $\epsilon_->0$ (Fig.~\ref{Es}c), and only Mu$^+_D$ for the case of Fig.~\ref{Es}d. (When $V$ or $V'\le 0$, only Mu$_D$ or Mu$_A$ are observed.)

It is clear from the left columns in Table \ref{Boffset} that Mu$^0$ with large hyperfine parameters are observed simultaneously with diamagnetic states in oxides with $E_g$ greater than $\sim$5 eV (except for BeO), which can be attributed to the Mu$^0_A$ and Mu$^+_D$ states that accompany $E^{-/0}$ and $E^{+/0}$, respectively (corresponding to Fig.~\ref{Es}b or \ref{Es}c).  Among those in which the calculated $E^{+/0}$ is in the band gap to predict Mu$^0_D$ states  ($\bigstar$ in the Table \ref{Boffset}), SiO$_2$, ZrO$_2$, and Y$_2$O$_3$ have indirect gaps smaller than $E_g$ around the $\Gamma$ point in the energy band structure, where the bottoms of the dispersive bands extend below (or near) $E^{+/0}$ \cite{Nekrashevich:14,Perevalov:06,Silva:12}.  The dashed parabolic curves found in Fig.~\ref{BA} symbolically represent the energetic extent of the band dispersion around the $\Gamma$ point. Thus, the diamagnetic states in these oxides are also interpreted as Mu$^+_D$. (In the actual experiment, there remains a possibility that the diamagnetic signal involves contributions from background and/or a fraction of muons that missed the electrons in the initial stage, although the Mu$^0$ can be attributed to one of two states attained in the case of $V$ or $V'\le0$.) Yet another exceptional case is $\alpha$-Al$_2$O$_3$ ($\blacklozenge$ in the Table \ref{Boffset}), for which the recent study suggests the occurrence of complex interactions between Mu, phonons, and excitons \cite{Vilao:21}.

The fact that Mu$^0$ is also observed at ambient temperatures in these materials is consistent with DFT calculations, because they predict that $E^{-/0}$ is situated far from the valence band ($\epsilon_-\gtrsim1$ eV);  the possibility for Mu$^0_A$ to change its valence state by capturing holes would be small even at ambient temperature.  Note that according to the DFT calculation, BeO corresponds to Fig.~\ref{Es}a, and the fact that only a single Mu$^0$ is experimentally observed suggests the situation that $V$ or $V'$ is negative.

Regarding the situation shown in Fig.~\ref{Es}c where $E^{-/0}$ is predicted to lie above the mid-gap, it is necessary to know the position of $E_{\rm CNL}$ for discussing whether $E^{-/0}$ can exist as an acceptor level within the band gap.  However, it has been argued that  $E_{\rm CNL}$ is almost identical with $E^{+/-}$ for H in binary compounds \cite{Walle:03} and in oxides \cite{Li:14,King:09a}, which has been in line with experimental implication for Mu in semiconductors with a relatively narrow band gap \cite{King:09b}.  Thus, it is reasonable to assume at this stage that the condition of $E_{\rm CNL}\simeq E^{+/-}\ge E^{-/0}$ holds for H/Mu in general. This can also be rephrased that hydrogen, with its ambipolarity, is a probe of the charge neutral level through its own charge state. 

In contrast to the case of wide-gap oxides, the predicted $E^{+/0}$ levels in those with a band gap of less than $\sim$5 eV fall in the conduction band ($\epsilon_+\le0$) without exception. 
Moreover, the $E^{-/0}$ level is deep in the gap (large value of $\epsilon_-$), merging to $E^{+/0}$ with decreasing $E_g$ (except for $\beta$-Ga$_2$O$_3$ and In$_2$O$_3$). Therefore, only the Mu$^+$ state is considered to be stable in these materials. However, there are a number of cases where the Mu$^0$ state, which is regarded as a shallow donor, is observed experimentally.  We will argue in the next section that these can be understood consistently by considering a common feature that the Mu$^0$ state exhibits polaron-like features, and by taking into account the strong electron-phonon coupling exhibited by the host. 

\subsection{Polaron states mimicking donor-like Mu}\label{pol} 
The electronic state of Mu$^0$ that behaves like a shallow donor (abbreviated as Mu$^0_S$, which is observed in oxides with $E_g\lesssim 5$ eV) is spatially extended, and its electronic structure can be discussed from the anisotropy of hyperfine interactions. As already mentioned in Section \ref{Mustat}, those in ZnO, TiO$_2$, and SrTiO$_3$ studied so far all exhibit an off-center polaronic character. The values of the hyperfine parameters in these materials are summarized in Table \ref{Acdip}, where they are expressed by applying Eq.~(\ref{Amu3}). In each case, $A_d$ is comparable to or greater than $A_c$, suggesting that the simple effective mass model of atomic H with a large Bohr radius is not suitable for understanding such electronic structures.

\begin{table}[b]
\begin{center}
\begin{tabular}{c|ccc}
\hline\hline
  & $A_c$ (MHz) & $A_d$ (MHz) & Refs. \\
 \hline
 \multirow{2}{*}{ZnO} & 0.579(9) & 0.177(5) & \multirow{2}{*}{\cite{Shimomura:02}}\\
 & 0.436(12) & 0.286(7) &\\
 TiO$_2$  & $-0.06(5)$ & $0.86(6)^*$ & \cite{Shimomura:15} \\
 SrTiO$_3$  & 1.4(3) & 15.5(2) &\cite{Ito:19} \\
 \hline
 GaN  & 0.079(22)& $0.258(22)$ & \cite{Shimomura:04} \\
\hline\hline
\multicolumn{4}{l}{\parbox{6.5cm}{\small ($^*$0.05$\mu_B$ on the nearest neighbor Ti)}}\\
\end{tabular}
\caption{Fermi contact term ($A_c$) and magnetic dipole interaction ($A_d$) in Mu$^0$ [see equation (\ref{Amu3})]. For ZnO, two states (Mu$_{1,2}$) have been observed from $\mu$SR experiments on single crystals, and the values for each are given. 
}\label{Acdip}
\end{center}
\end{table}

Interestingly, these electronic states show distinct similarities to the off-centered STE in alkali halides consisting of the hole localized on a halogen dimer ($X_2^-$, known as V$_{\rm k}$ centers) and the electron at the halogen vacancy (corresponding to the F center) situated next to the $X_2^-$ dimer \cite{Williams:86,Song:96,Kanno:97,Itoh:97}.
If we consider OMu$^-$ as an analogue of the hole-localized halogen dimer, then the electron attracted to it is presumed to avoid the Coulomb repulsion from the nearby anions, being localized at the cation.   Thus, the electronic structure of Mu$^0_S$ can be interpreted as a result of compromise among the strong electron-phonon coupling that promotes the localization of electrons, the Coulomb attraction from Mu$^+$, and the Coulomb repulsion from the neighboring O$^{2-}$.  In other words, Mu$^0_S$ is an STE-like state involving Mu, mimicking the shallow donor Mu$^0_D$ state. 

This similarity suggests that the Mu-exciton interaction discussed at the beginning also contribute to the formation of Mu$^0_S$, while the local charge polarity of OMu$^-$ (= Mu$^+_D$) is opposite to the interstitial Mu$^+_A$. In fact, excitons bound to various donor/acceptor impurities are known to exist in ZnO and TiO$_2$, and their local electronic structure has been investigated by photoluminescence spectroscopy \cite{Meyer:04,Meyer:07,Gallart:18}. A very recent report provides a variety of H-related bound excitons in ZnO \cite{Heinhold:17}, although that corresponding to the bond-center H$^0$ seems missing (probably due to the small yield). Therefore, the polaron-like bound states observed in these materials can be qualitatively understood, including the reason for the off-centered electronic structure, by adopting a reversed viewpoint that OMu$^-$/OH$^-$ serves as an electron trap to form a complex state analogous to the donor-bound exciton  \cite{FunSem}. 

We have long been plagued by the problem that the $E^{+/-}$ levels obtained by DFT calculations for these oxides not only contradict the observation of neutral Mu$^0$ states, but also do not correspond to ``shallow'' donor levels.  Our new model provides a fundamental shift from such a naive interpretation in that the shallow donor-like Mu$^0$ corresponds to a relaxed-excited state, where the bound state is formed by the capture of the exciton electron to the $E^{+/0}$ level with the help of strong electron-phonon interactions.

The localization of electrons is promoted by the electron-phonon coupling, which has been inferred to be strong in ZnO  \cite{Yang:09}, TiO$_2$  \cite{Yagi:96}, and SrTiO$_3$ \cite{Spinelli:01} from the suppression of electron mobility at higher temperatures. In contrast to these cases (including ZnO), a monotonous enhancement of carrier mobility is observed with increasing temperature in the pristine a-IGZO \cite{Ide:19}, in which no shallow state is observed for Mu localized near the Zn-O bond center \cite{Kojima:19}.  Thus, considering that the local atomic configuration of Mu/H is almost identical between ZnO and IGZO \cite{Kojima:19}, the strong electron-phonon coupling is another important factor to enhance the localization of electrons around the OMu$^-$ complex. This invokes the precaution that the activation energy for the promotion from Mu$_D^0$ to Mu$_D^+$ cannot be simply attributed to $\epsilon_+$; it includes the contribution of the adiabatic potential barrier for the electron between the localized state and the continuous state \cite{Ito:19}. (A similar argument would be the case for $\alpha$-Al$_2$O$_3$ in the previous section.)  The effect of the electron-phonon coupling has also been theoretically investigated in terms of the duality of the conduction band carriers in TiO$_2$ and SrTiO$_3$, which can be in both the continuous state and the deep level (self-trapped) state in the gap \cite{Janotti:13}. 

In any case, when the donor-like Mu$^0$ (Mu$_D^0$) is observed experimentally despite that the DFT calculation predicts $E^{+/0}$ is in the conduction band ($\epsilon_+\lesssim0$), the possibility of electron localization at Mu$_D^+$ enhanced by the electron-phonon coupling is plausible as the primary cause. It should be pointed out that this is also likely for the ``shallow donor Mu$^0$'' observed in GaN \cite{Shimomura:04}, for example (bottom row of Table \ref{Boffset}). Previous theoretical calculations have predicted that $\epsilon_+\sim0$, and that interstitial H in GaN does not form shallow donor levels as described by the effective mass model \cite{Neugebauer:95}. (That the electronic state of Mu was discussed in relation to $E^{+/-}$ which was located deep in the band gap was also the problem.)  In contrast, the reported hyperfine interaction has $c$-axis symmetry with $A_\parallel=0.337(10)$ MHz and $A_\perp=-0.243(30)$ MHz. Deriving the parameters of Eq.~(\ref{Amu3}) from these values gives the values shown in Table \ref{Acdip}, indicating that the magnetic dipole interaction is dominant, as in the case of TiO$_2$. The presence of strong electron-phonon coupling has been inferred from the suppression of the hole mobility in GaN at high temperatures \cite{Nakamura:92}. Furthermore, the occurrence of H-exciton interaction has been also demonstrated by photoluminescence spectroscopy \cite{Chtchekine:99}. These situations suggest that Mu$^0$ in GaN can be also understood as analogous to a donor-bound exciton in which an electron is localized near Mu$_D^+$ by electron-phonon coupling to mimic the shallow donor state.

Now, provided that the bound-exciton model properly explains the Mu$^0_S$ state, it is expected that the localization of electrons in the vicinity of Mu will be determined by stochastic processes which may be also influenced by the presence of other impurities and defects nearby. Therefore, the diamagnetic states in these oxides can be attributed to the Mu$_D^+$ state predicted by the DFT calculation, where the relative yield between Mu$^0_S$ and Mu$_D^+$ is presumably determined by the local carrier (exciton) density. 

At this stage, it may be worth a comment on the claim for the existence of Mu$^0$ with shallow levels in SnO$_2$ \cite{Cox:06b,King:09b}. The reported hyperfine parameter is as small as $A/2 = 0.045(1)$ MHz with small relative yield ($\sim$3.6\%) \cite{King:09b}, and it is not clear how it was distinguished from slow depolarization due to nuclear magnetic moments ($\Delta\simeq0.03$ MHz \cite{Cox:06b}) and its apparent reduction due to muon diffusion (that mimics the promotion of Mu$^0$ to Mu$^+$). In any case, the important point is that the majority of implanted Mu is in a diamagnetic state, which can be attributed the donor-like Mu$^+$ state. 

Recently, a detailed comparison between the magnitude of the internal magnetic field at the Mu position observed in the magnetically ordered phase of Cr$_2$O$_3$ and Fe$_2$O$_3$ (which contain magnetic ions) and the local electronic state predicted by DFT calculations shows that the electrons associated with Mu are localized in the $3d$-orbitals of the Cr/Fe ions \cite{Dehn:20,Dehn:21}. Since the electron-phonon coupling is known to be strong in these materials, these off-centered electronic states indicate that a similar mechanism is at work in the formation of polarons.

\subsection{Fast diffusion of acceptor-like Mu}\label{Mudiff}
\begin{figure}[t]
	\centering
	\includegraphics[width=0.9\linewidth,clip]{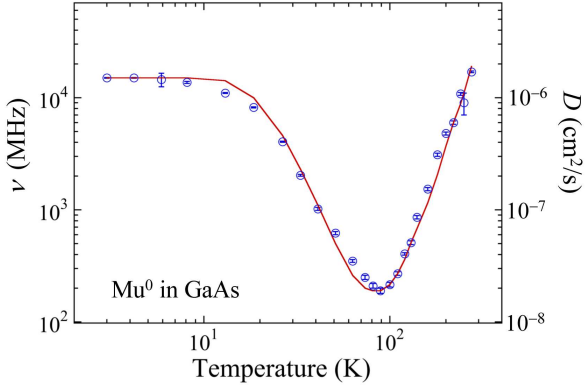}
	\caption{Temperature dependence of Mu$_A^0$ hopping frequency ($\nu$) observed in GaAs, where the solid curve represents theoretical prediction by the quantum diffusion model (after Ref.\cite{Kadono:91}). The corresponding diffusion constant ($D$) is derived from the relationship $D=a^2\nu/32$ (with $a=0.56535$ nm being the lattice constant). }
	\label{mudif}
\end{figure}

It is expected that the donor-like Mu in non-metallic compounds (Mu$_D$) forms a covalent bond ($\sigma$ bond) with anions and requires relatively large activation energy to migrate between the equivalent sites, whereas many examples of fast diffusive motion of Mu$^0$ have been reported. In elemental semiconductors (Si, Ge), macroscopic diffusion coefficients have been inferred for Mu$^0_A$ (the ``normal muonium'') occupying the center of the tetrahedral interstitial position (the $T_d$ site) \cite{Lichti:08} through the diffusion-limited trapping of Mu$^0$ to extrinsic impurities \cite{Patterson:88}. The detailed temperature dependence of the hopping frequency of Mu$^0$ has been revealed via the dynamical fluctuations of the nuclear hyperfine interaction (i.e., that between nuclear spins and an orbital electron of Mu$^0$ \cite{Patterson:88}). Figure~\ref{mudif} shows an example in GaAs \cite{Kadono:91,Kadono:94}, where the relevant Mu$^0$ state was later inferred to be located at the center of the cationic Ga tetrahedral cage  (i.e., Mu$^0_A$) \cite{Lichti:01}. Fast hopping motion of Mu$^0$ at the $T_d$ site was also reported in diamond (C), where a qualitatively similar temperature dependence was observed \cite{Gxawu:05}. Moreover,  such a behavior is also suggested for the Mu$^0_A$-like states in various compounds other than oxides (including iron pyrite, see the next section). These observations suggest that the acceptor-like Mu$^0$ have an intrinsic property of fast diffusion in common. Here, it must be emphasized that Mu$^0$ in the presence of high density carriers behave as if it is in the diamagnetic state due to the motional narrowing of the hyperfine interaction caused by the fast spin/charge exchange \cite{Chow:00}.  In such a situation, Mu diffusion can be monitored via the fluctuation of local magnetic fields exerted from nuclear magnetic moments, where the muon depolarization is described by the dynamical Gaussian Kubo-Toyabe function \cite{Hayano:79}.

In general, Mu/H in solid crystals is presumed to form a polaronic state in itself due to the interaction with the lattice (phonons), and the self-diffusion is well understood by the hopping motion of small polarons \cite{Holstein:59b}. From the previous discussion on the electronic state of H$^0_i$ in alkali halides, the antibonding character of acceptor-like Mu/H with surrounding anions emerges as an additional factor to enhance the isolation of Mu$_A^0$.  Because of the light mass, the tunneling effect is known to dominate the hopping frequency of Mu/H even at ambient temperatures. The high symmetry of the local structure around Mu$^0_A$ stemming from the antibonding character seems to be advantageous for having a relatively large tunneling matrix. 

More specifically, there are two modes of tunneling motion: one is the thermally activated tunneling, in which thermal excitations (phonons) enhances tunneling by compensating the self-trapping energy (phonon-assisted tunneling), and another in which phonons inhibit the intrinsic (coherent) tunneling by blurring the final state \cite{Kagan:92}. As shown in Fig.~\ref{mudif}, hopping frequency increases with increasing temperature on the high temperature side ($T\gtrsim\varTheta_D/3$, with $\varTheta_D\simeq347$ K being the Debye temperature of GaAs) as the phonon-assisted tunneling becomes dominant. Meanwhile, the hopping frequency increases again at low temperatures as the phonon scattering decreases with decreasing temperature, and converges to a frequency determined by the magnitude of the tunneling matrix.
This behavior was first observed clearly for Mu$^0$ in alkali halides \cite{Kiefl:89,Kadono:90}, and is expected to be true for oxides as well, since similar situations are suggested for elemental semiconductors and group 13-15 compounds.

In fact, it is known that the atomic Mu$^0$ is emitted into the vacuum when a silica powder sample (SiO$_2$) placed in a vacuum is irradiated with muons \cite{Marshall:78,Beer:86}, indicating that the Mu$^0$ produced within SiO$_2$ particles (sub-micrometers in diameter) diffuses rapidly to reach the crystalline surface. In addition, motional narrowing due to fast diffusion of Mu$^0$ in SiO$_2$  above 180 K has been pointed out as the reason for the disappearance of the anisotropy in the hyperfine parameters associated with the electric quadrupole interaction of Mu$^0$. Conversely, this is consistent with the interpretation of Mu$^0$ in SiO$_2$ as an acceptor-like state (not bound to O) in the previous section. 
Such fast diffusion of acceptor-like Mu/H may have some implications for recent studies of hydride conduction in metal oxyhydrides \cite{Kobayashi:16,Fukui:21}.

\subsection{Comparison with hydrogen in thermal equilibrium}\label{ImpH}
In contrast to the case of Mu, the electrical activity of H in a material of equilibrium state is considered to be determined by the position of the $E^{+/-}$ level in the energy band structure. When $E^{+/-}>E_c$, the electronic level strongly hybridizes with the conduction band and falls to the bottom of the band, and H becomes an electron donor regardless of temperature. The same is true for the electric activity as acceptors, which is determined by the relationship between $E^{+/-}$ and $E_v$. In addition, when $E^{+/-}$ is in the midgap ($E_v<E^{+/-}<E_c$), the hybridization is weak and the isolated H is electrically inactive. Therefore, the behavior of $E^{+/-}$, including whether or not it shows any regularity, is very important in predicting the electrical activity of H. In fact, $E^{+/-}$ in group 13-15 compound semiconductors and oxides has been vigorously evaluated by first-principles calculations in the past decades.

As mentioned earlier, it was suggested from the earlier DFT calculations that $E^{+/-}-E_{\rm vac}$ takes a universal value common to oxides \cite{Kilic:02}.
This is in stark contrast to the cases of $E_c$ and $E_v$ vs $E_{\rm vac}$ that vary among oxides without systematics. Such an alignment of $E^{+/-}$ was originally proposed for deep impurity levels in binary compound semiconductors, and it has been suggested that the local impurity-host interaction (e.g., strength of covalency) determines the depth for each class of compounds \cite{Caldas:84}. This alignment hypothesis has been applied to a series of oxides \cite{Schmickler:86}, from which the candidates with shallow donor levels were proposed \cite{Kilic:02}.

\begin{figure}[b]
	\centering
	\includegraphics[width=\linewidth,clip]{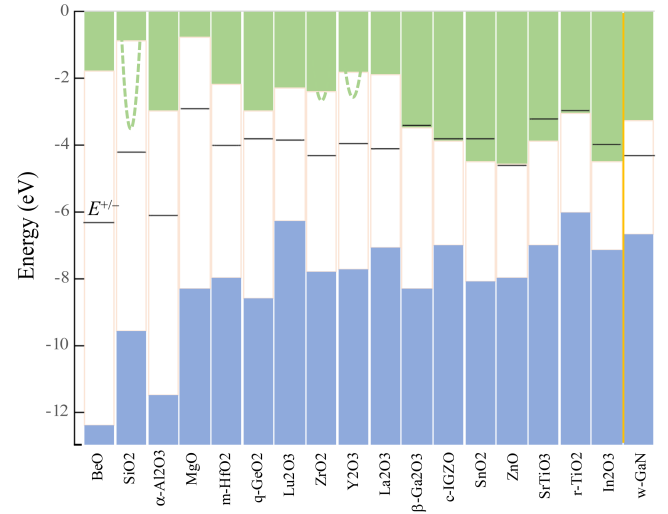}
	\caption{The equilibrium charge-transition levels ($E^{+/-}$) in Table \ref{Boffset} plotted on the band structure aligned to the vacuum level. }
	\label{BAc}
\end{figure}

A model was also proposed in which the alignment of $E^{+/-}$ occurs midway between the $\sigma^*_D$ and $\sigma_A$ bonding levels in Fig.~\ref{mue}a (the antibonding level raised above the conduction band by the OH bond and the O $2p$ level generated by the metal-hydrogen bond, respectively) \cite{Peacock:03,Xiong:07}. Thus $E^{+/-}$ is expected to be located near the charge neutral level ($E_{\rm CNL}-E_{\rm vac}\simeq-5$ eV).  Assuming that $E_{\rm CNL}$ is at a constant offset to the top of the valence band (which mainly consists of O 2$p$ orbitals), the H levels are expected to exhibit the tendency of alignment around $E_{\rm CNL}$.  This model also appears to account for deeper levels such as those in SiO$_2$, Al$_2$O$_3$, and MgO, which were not well explained by the previous model of offset alignment from $E_{\rm vac}$.   The $E^{+/-}$ levels relative to $E_{\rm vac}$ estimated by DFT calculations quoted in Table \ref{Boffset} are shown in Fig.~\ref{BAc}, which are qualitatively in line with these scenarios.  

These predictions about $E^{+/-}$ have aroused a great deal of interest, and the results of experimental verification studies conducted up to 2005 with special focus on shallow donor Mu levels in oxides have been reported in two review papers \cite{Cox:06a,Cox:06b}. According to them, SnO$_2$, La$_2$O$_3$, TiO$_2$ (rutile) exhibited shallow donor levels, and Al$_2$O$_ 3$, MgO, and SiO$_2$ seemed to have deep in-gap levels, which were both in line with the predictions. Meanwhile, Bi$_2$O$_3$, HgO, and Sb$_2$O$_3$ turned out to be characterized by large hyperfine parameters suggesting deep levels in contradiction to the predicted shallow donor state. 

As has been already discussed in Sect.~\ref{muh}, such discrepancies can be resolved by departing from the implicit assumption that the electrical activity of Mu is dominated by $E^{+/-}$ in analogy with H. (The partial success in the previous attempts can be attributed to the fact that the difference between $E^{+/-}$ and $E^{+/0}$ was relatively small in some oxides.) 
The electronic states of Mu in elemental (group 14) and 13-15 compound semiconductors have come to a coherent understanding by assuming that it varies with the occupancy of the $E^{\pm/0}$ levels. This also allowed to estimate $E^{+/-}$ by interpolation from $E^{\pm/0}$, which was compared with the prediction of DFT calculations in parallel with the case of H  \cite{Lichti:08}.  

For the case of oxides, the donor-like Mu$_D$ is in the diamagnetic state (Mu$^+_D$) in most cases, for which it is difficult to estimate $\epsilon_+$ ($<0$) by observing change in the valence state with elevating temperature (i.e., the ionization of Mu$_D^0$).  In contrast, Mu$_A$ is often observed in a paramagnetic state (Mu$^0_A$).  Therefore, $\epsilon_-$ can be evaluated experimentally by observing the promotion of Mu$^0_A$ to Mu$^-_A$ at sufficiently high temperatures ($kT\gg\epsilon_-$). It is a future task to evaluate the pinning level $E^{+/-}$ by combining the information about $\epsilon_-$ and those for the Mu$_D$ states, and to discuss what kind of regularity can be presumed for $E^{+/-}$ in oxides.

\subsection{Implications to Mu/H in other ionic compounds}\label{Impl}
The model of the bistable relaxed-excited states for Mu (Mu$_D$ and Mu$_A$) is useful for understanding electronic states of Mu in other materials that were previously thought to be isolated anomalies. In the following, we will discuss Mu/H in iron pyrite (FeS$_2$) and sodium alanate (NaAlH$_4$) as such examples, which can be important materials for green technologies (solar cells, hydrogen storages) aimed at a decarbonized society.  Some general remarks are also made on the oxides that have been the focus of $\mu$SR studies over decades from the viewpoint of strongly correlated electron systems.\\
 
\subsubsection{\it FeS$_2$}
Iron pyrite has long been attracting attention as a promising optoelectronic material due to its suitable indirect band gap ($E_{\rm g}\simeq0.95$ eV) and high absorption coefficient ($>10^5$ cm$^{-1}$ at $E_{\rm g} \pm0.1$ eV), which opens up great potential for emerging renewable energy applications, including photovoltaics, photodetectors, and photoelectrochemical cells. Until recently, the low open-circuit photovoltage ($V_{\rm oc}$$\sim$0.2 V) that comprises the main obstacle to these applications have been attributed to surface defect states. Regarding the heterogeneous bandgap and Fermi level pinning, sulfur vacancies has been a primary suspect. However, recent theoretical investigations suggest that this is not necessarily the case \cite{Yu:11}, leading to the renewed interest to the circumstantial evidence that H is involved in the $n$-type conductivity of unknown origin with activation energies less than 0.01 eV \cite{Ennaoui:93}.  Moreover, electrochemical experiment suggests strikingly fast H diffusion in pyrite (corresponding diffusion coefficient $D_H\ge2\times10^{-6}$ cm$^2$/s, comparable to that in bcc metals at ambient condition) \cite{Wilhelm:83,Bungs:97}, which is further enhanced after saturation of defects by H \cite{Bungs:97}.

Our recent muon study has serendipitously revealed that there are two different Mu states in pyrite, one is situated near the center of an Fe$^{2+}$-cornered tetrahedron with a nearly isotropic hyperfine parameter [Mu$^0_{\rm p}$, $A=0.41(4)$ GHz], and the other as a diamagnetic state (Mu$_{\rm d}$) located near the antibonding site of the S$_2^{2-}$ dimer \cite{Okabe:18}. Their response to thermal agitation indicates that the Mu$_{\rm d}$ center accompanies the $E^{+/0}$ level within the conduction band, while the electronic structure of Mu$^0_{\rm p}$ is more isolated from the host than Mu$_{\rm d}$ to form a deep in-gap level.  The possibility of fast migration for Mu$^0_{\rm p}$ to form a complex defect state with existing impurities at low temperatures ($T\le100$ K) and its fast diffusion at higher temperatures upon release from the complex ($T\ge150$ K) has been suggested  \cite{Okabe:18}.

\begin{figure}[t]
	\centering
	\includegraphics[width=0.95\linewidth,clip]{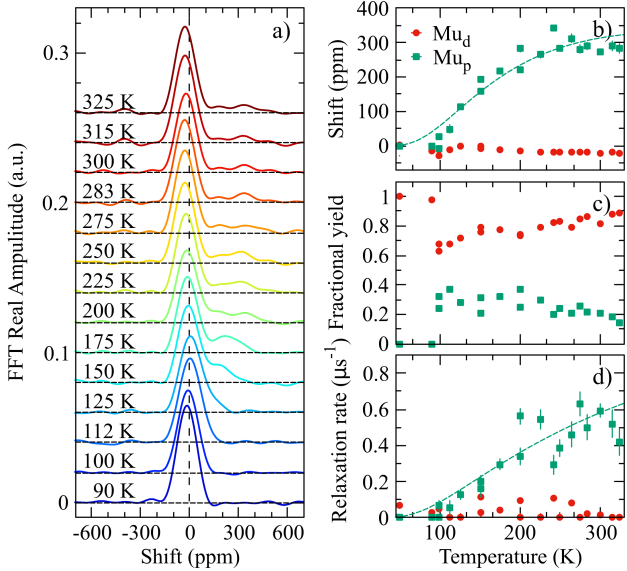}
	\caption{Temperature dependence of the muon Knight shift observed in FeS$_2$. a) Fast Fourier transform of the $\mu$SR spectra at typical temperatures measured under an external field of 6 T. The results of curve-fit analysis are shown for b) the frequency shift, c) fractional yields, and d) relaxation rate vs. temperature ($T$). For the dashed curves in b) and d), see text.}
	\label{fes2}
\end{figure}

That the Mu$^0_{\rm p}$ state is also present at higher temperatures is confirmed by our very recent additional muon Knight shift measurements. As shown in Fig.~\ref{fes2}, Mu in pyrite has a component with a large positive shift that cannot be explained by chemical shifts associated with Mu$_{\rm d}$, indicating that this is the signal from Mu$^0_{\rm p}$. Furthermore, the lack of the  low-frequency side signal normally observed for Mu$^0$ is readily understood by considering that its electronic state is subject to a fast spin/charge exchange reaction with carriers \cite{Chow:00} (as the sample is unintentionally  $n$-type doped). The tendency of the shift to have a maximum around 300 K and a decrease at higher temperatures is also consistent with the expected Curie law dependence on temperature. The dashed curves in Fig.~\ref{fes2} are obtained by the least-square fit using a product of the Curie law and Arrhenius law, $K\propto(C/T)\exp(-E_a/kT)$ for b) with $E_a=36(3)$ meV, and using the Arrhenius law, $\lambda\propto\exp(-E_p/kT)$ for d) with $E_p=25$ meV   (which are in reasonable agreement with the values estimated from the $T$ dependence of the longitudinal depolarization rate \cite{Okabe:18}). 

In retrospect, these two states can be readily understood as analogues of the donor/acceptor-like Mu states in oxides; Mu$_{\rm d}$ is assigned to the Mu$_D$ forming a complex with S$_2^{2-}$ anion whose electronic level falls in the conduction band to leave persistent Mu$_D^+$ state, and Mu$^0_{\rm p}$ is to the Mu$_A$ forming a complex state with four Fe$^{2+}$ cations (which was previously speculated as a deep donor-like state of unknown origin \cite{Okabe:18}). It is noteworthy that the latter also exhibits the characteristic fast diffusion, which is also shared by the case of H in pyrite.    Such similarities/correspondences suggest that the behavior of Mu/H in other metal chalcogenides is also understood within the same framework discussed for oxides.\\

\subsubsection{\it NaAlH$_4$} 
Sodium alanate belongs to a class of complex hydrides that has emerged as one of the most promising hydrogen storage materials in recent years.   While these materials commonly have an advantage of high H weight percentage, they are generally characterized by extremely slow H cycling kinetics that has precluded applications.  The drastic improvement on the kinetics in NaAlH$_4$ at moderate temperatures ($\sim$100 C$^\circ$) attained by transition metal doping (e.g. a few percent of Ti) \cite{Bogdanovic:97} has triggered a surge of investigation for the microscopic mechanism of the enhancement by various local probes including implanted Mu.  Here, the muon study aims to mimic the behavior of interstitial H which is relevant for understanding the intermediate (non-equilibrium) states of the kinetic processes inside the compound during H intake/release.  

The NaAlH$_4$ crystal consists of Na$^+$ cations and alanate (AlH$_4^-$) anions in the tetragonal structure (space group $I4_1/a$). It is inferred from DFT calculations that the band structure is that of an ionic insulator ($E_g\simeq6.7$ eV), in which the valence band mainly consists of hydrogen $s$-$p$ orbitals from alanate ions while the conduction band is dominated by those from Na$^+$ ions \cite{Setten:07}.  In our previous report, the implanted Mu exhibited two different states, where the $\mu$SR signal with spontaneous spin precessions characteristic of a three-spin system consisting of $\mu^+$-two proton nuclear spins was attributed to the AlH$_4^-$-Mu$^+$-AlH$_4^-$ complex, and the signal indicating Gaussian Kubo-Toyabe relaxation was attributed to an isolated Mu$^+$ at the octahedral interstitial position (O-site)  \cite{Kadono:08}. By applying our model, we are now able to provide a more detailed discussion on the local electronic structure of these observed states.

\begin{figure}[t]
	\centering
	\includegraphics[width=0.95\linewidth,clip]{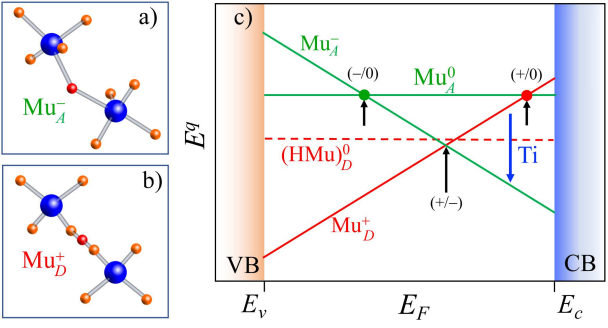}
	\caption{The proposed models of a) Mu$^-_A$ where a H$^-$  atom bridging two AlH$_4^-$ units in the AlH$_4^-$-H$^-$-AlH$_4^-$ complex \cite{Short:09} is replaced by Mu$^-$, and b)  Mu$^+_D$ where a Mu$^+$ atom bridges two H atoms of AlH$_4^-$ by the hydrogen bonding \cite{Kadono:08}. c) Schematic diagram showing the formation energy of Mu-related defects ($E^q$) vs.~Fermi level ($E_F$) in NaAlH$_4$. The initial Mu states correspond to those accompanying $E^{\pm/0}$ levels. The vertical arrow indicates the redox process catalyzed by Ti doping for $E_F>E^{+/-}$.}
	\label{sodal}
\end{figure}

From first-principles DFT calculations on the H-related native defects, the state corresponding to H$^-_i$ is predicted to form the AlH$_4^-$-H$^-$-AlH$_4^-$ complex, where H$^-$ is stabilized by the bonding character with Al$^{3+}$. While for H$^+_i$, it reacts with AlH$_4^-$ to generate the H$_2^+$ molecule (2AlH$_4^-+$H$^+\rightarrow$AlH$_3^-\cdot$AlH$_4^-$ + H$_2^+$) \cite{Peles:07,Lodziana:08,Short:09}. It is tempting to consider that the state observed as a few-spin system may correspond to H$^-_i$. However, as shown in Fig.~\ref{sodal}a, there are six proton nuclei on two nearest neighboring AlH$_4$'s equidistant from Mu that will lead to the Gaussian Kubo-Toyabe relaxation (the contribution of $^{23}$Na nuclei is negligible). Thus, the state previously attributed to the isolated Mu$^+$ at the O site is interpreted to be the AlH$_4^-$-Mu$^-$-AlH$_4^-$ complex, corresponding to M$_A^-$. In fact, the observed Gaussian relaxation rate ($\Delta=0.4$--0.5 MHz) was significantly greater than that expected for the O site (0.33 MHz), suggesting the displacement of Mu towards AlH$_4^-$ units \cite{Kadono:08}.

The DFT calculations also suggest that the H$_i^-$ ions can move with relatively small activation energy (0.16 eV) \cite{Short:09}. The muon study has also shown that the Mu$_A^-$-like state exhibits diffusion by hopping motion with a rate comparable with that observed for Mu$^+$ in fcc metals ($\sim$10$^{-9}$-10$^{-10}$ cm$^2$/s at ambient temperature), and that its relative yield is significantly enhanced in place of the Mu$_D^+$ state by the Ti doping \cite{Kadono:08}. That the hopping rate of Mu$_A^-$ is much smaller than that of Mu$_A^0$ (by 10$^{-3}$, see Fig.~\ref{mudif}) may be attributed to the bonding character between the H 1$s$ and Al $sp^3$ orbitals.

Regarding Mu$^+_i$, when it forms a diatomic HMu$^+$ molecule by picking up a H atom (as in parallel with the case of native defects), it is expected to behave as a paramagnetic center (``muonated radical'', where muon spin is subject to a large hyperfine field from an unpaired electron).  But no such paramagnetic Mu has been observed experimentally. Moreover, even if it forms a neutral HMu molecule (which would also exhibit spontaneous muon spin precession characteristic to two-spin systems), the distance between Mu and proton nuclear spin ($r=0.074$ nm) is too small to account for the experimentally deduced value (= 0.145 nm; note that the spin precession frequency is inversely proportional to $r^3$). The scarcity of the isolated interstitial H for Mu$^+$ to encounter in the ambient condition also disfavor the HMu molecular formation. Therefore, it is reasonable to assume that the observed signal is not from the HMu molecule (i.e., the thermodynamical ground state) but from the AlH$_4^-$-Mu$^+$-AlH$_4^-$ complex (see Fig.~\ref{sodal}b) \cite{Kadono:08}, where the latter is realized as a metastable state (or stabilized by isotope effects due to the light mass of Mu).

The relationship between the formation energy and Fermi level  of these states based on the results of DFT calculations is depicted in Fig.~\ref{sodal}c \cite{Peles:07,Short:09}, where $E^+$ for Mu$^+_A$ is presumed to follow that of the H vacancy (V$_{\rm H}^+$). From the viewpoint of the local electronic excitation [e.g., Eq.(\ref{H0i})], the absence of the Mu$^0$ state suggests that the (AlH$_3)^-$ state is not stable as an excited hole in the anion sublattice, and that the electronic excitation associated with muon implantation is dominated by the process AlH$_4^-+h\nu\rightarrow$(AlH$_3)^0$ (= V$_{\rm H}^+$) $+$H$^-_i$. For Mu$_D^+$, it is natural to assume the occurrence of the AlH$_4^-$-Mu$^+$-AlH$_4^-$ complex as a metastable state. The relative yields of Mu$^+_D$ and Mu$^-_A$ are consistent with the presumption that these states are generated with equal probabilities to satisfy the charge-neutrality condition. 

Given the formation energy shown in Fig.~\ref{sodal}, we are now more confident that the role of Ti is to convert the Mu$_D^+$ state to the mobile Mu$_A^-$ state; ${\rm Mu}_D^++2e^-\rightarrow{\rm Mu}_A^-$, which can be interpreted as a redox process catalyzed by Ti (indicated by the vertical arrow in Fig.~\ref{sodal}).  In the thermal equilibrium, it is likely that $E_F$ increases by Ti doping (electron donation), which reduces $E^-$ and promotes H$_A^-$ formation against H$_D^+$. \\

\subsubsection{\it Other narrow-gap oxides}
There is a massive body of $\mu$SR research on transition metal oxides exhibiting superconductivity and/or magnetism, where implanted Mu has been implicitly presumed as a bystander (bare $\mu^+$) with minimal perturbations to the host in most cases. As shown in Table \ref{tmo} for typical compounds, these materials often have a small band gap (1--2 eV) even in the insulating phase, for which Mu is observed in a diamagnetic state with the relative yield of nearly 100\%. Their band structure aligned to the vacuum level is also shown in Fig.~\ref{BA}, where the bottom of conduction bands lies between $-3$ and $-5$ eV. These suggest a situation similar to Fig.~\ref{Es}d where the donor levels for Mu are in the conduction band to allow only for the Mu$^+$ state.  Moreover, most of the muon studies for cuprates is concerned with their metallic phases attained by carrier doping, where implanted muons have virtually no chance of forming the Mu$^0$ state due to the electrostatic shielding of the positive charge by conduction electrons. Thus, it is concluded that the interpretation of $\mu$SR results in those compounds are virtually unaffected by the ambipolar property of Mu in most cases.

However, it should be remembered for insulating magnetic compounds that implanted Mu can attract an electron to form a polaronic state under certain conditions, which makes the interpretation of muon result complicated by affecting the local magnetic structure \cite{Dehn:20,Dehn:21}.

\begin{table}[t]
\begin{tabular}{c|ccc}
\hline\hline
Mater. & $E_g$ (eV) & Mu & Refs.\\
\hline
La$_2$CuO$_4$ 		& 1.5--2 & Mu$^+$  & \cite{Budnick:87} \\
YBa$_2$Cu$_3$O$_6$ 	& $\sim$1.5 & Mu$^+$ & \cite{Nishida:87} \\
LaMnO$_3$ 		& $\sim$1.5 & Mu$^+$ & \cite{Heffner:01} \\
Ne$_2$CuO$_4$ 		& 1--1.5 & Mu$^+$ & \cite{Luke:90} \\
\hline
\end{tabular}
\caption{The electronic state of Mu in cuprates and manganites, where the bandgap of the parent compounds are listed \cite{Yunoki:07}.} \label{tmo}
\end{table}

\section{Conclusion}
In this paper, we have introduced a model that explain the behavior of Mu and H in insulating oxides in a coherent manner.  For the Mu part, it is crucial to consider non-equilibrium effects for the construction of a viable model. This leads to the conclusion that the information obtained from Mu is not about the equilibrium charge-transition levels ($E^{+/-}$), but about the donor/acceptor levels ($E^{\pm/0}$). The conclusion has been supported by the consistency between the electronic states of Mu predicted by the position of $E^{\pm/0}$ levels in the band structure evaluated by first-principles DFT calculations for H and those experimentally observed in oxides. In addition, by establishing a model that allows such a systematic understanding, two findings have been inductively derived. The first is the suggestion that the Mu$^+_D$-bound excitons are the origin of ``shallow donor''-like Mu states in relatively narrow-gapped oxides ($\lesssim5$ eV), as inferred from their STE-like electronic structure. This will bring about a major change of perspective in our understanding of the electronic structure of Mu. Second, it is revealed that the acceptor-like Mu$^0$ has a common property of fast diffusion in oxides. This would be qualitatively true for H as well, although there is a correction for isotope effects. 

These findings indicate that, by integrating i) the experimental data on the $E^{\pm/0}$ levels obtained from the Mu studies, ii) those on the $E^{+/-}$ levels in the thermal equilibrium of H, and iii) the first-principles calculations based on the DFT theory, a coherent understanding on the behavior of hydrogen in materials can be achieved.  The model will be also useful for clarifying the role of hydrogen in a wide range of non-metallic materials. 

\begin{acknowledgements}
The works of the authors quoted in this paper were conducted in collaboration with many colleagues. We would like to appreciate helpful discussions with K. Asakura, K. Fukutani, K. Ide, S. Iimura, T. U. Ito, W. Higemoto, Y. Kamiya, R. F. Kiefl, K. M. Kojima, R. L. Lichti, W. A. MacFarlane, S. Matsuishi, H. Miwa, F. Oba, N. Ohashi, T. Ohsawa, T. Prokschar, J. Robertson, M. Saito, K. Shimomura, A. L. Shluger, A. Suter, S. Tsuneyuki, and R. Vil\~ao. This work was supported by the MEXT Elements Strategy Initiative to Form Core Research Center for Electron Materials (Grant No. JPMXP0112101001) and JSPS KAKENHI (Grant Nos. 19K15033 and 17H06153).
\end{acknowledgements}

\vspace{2cm}

\section*{APPENDIX A: \msr\ spectrum without unpaired electrons}
In the following, we summarize the typical cases of the hyperfine interactions between muon and nuclear/electron spins that is represented by the effective local magnetic field  ${\bm H}({\bm r})$ and the corresponding time variation of the muon spin polarization (time spectrum) to be observed.

Let us first consider non-magnetic materials where there are no unpaired electrons (correspond to Fig.~\ref{muc}b, f, h, and i). In these cases, the origin of ${\bm H}({\bm r})$ is none other than the nuclear magnetic moments of the host. In general, the term ``hyperfine interaction" includes both magnetic dipole interaction and the Fermi contact interaction. However, since both nuclei and muons are well localized in the ground state, the interaction between them is mainly magnetic dipole interaction. The Hamiltonian is then given as
\begin{align}
{\cal H}/\hbar =  {\cal H}_{\rm Z}/\hbar+{\cal H}_{\rm d}/\hbar, \\
{\cal H}_{\rm d}/\hbar =\gamma_\mu\gamma_I{\bm S}_\mu\sum_i\hat{A}_{\rm d}^i{\bm I}_i,\label{nucldip}
\end{align}
where ${\cal H}_{\rm Z}$ represents the Zeeman interaction for muon and nuclear spins, ${\bm S}_\mu$ is the muon spin, $\gamma_\mu=2\pi \times 135.53$ MHz/T is the gyromagnetic ratio of muon spin, ${\bm I}_i$ is the nuclear spin at distance $r_i$ on the $i$th lattice point, $\gamma_I$ is the gyromagnetic ratio of the nuclear spin, and $\hat{A}_{\rm d}^i$ is the magnetic dipole tensor
\begin{equation} 
(\hat{A}_{\rm d}^i)^{\alpha\beta}
=\frac{1}{r^3_i}(\frac{3\alpha_i \beta_i}{r^2_i}-\delta_{\alpha\beta}) \:\:(\alpha,\beta=x,y,z),\label{diptensor}
\end{equation} representing the hyperfine interaction between muon-nuclear magnetic moments. In the case of zero-external field (${\cal H}_{\rm Z}=0$), the effective magnetic field expressed as
\begin{equation}
{\bm H}({\bm r})={\bm H}_{\rm d}=\gamma_I\sum_i\hat{A}_{\rm d}^i\overline{{\bm I}}_i\label{bdip_n}
\end{equation}
 is used to obtain the effective Hamiltonian
\begin{equation}
{\cal H}/\hbar =\gamma_\mu{\bm S}_\mu\cdot{\bm H}_{\rm d},
\end{equation}
and the time evolution of the muon spin polarization ${\bm P}(t) =\langle {\bm S}_\mu(0)\cdot{\bm S}_\mu(t)\rangle/|S_\mu^2|$ can be obtained analytically using the density matrix of the muon-nucleus spin system for a small number of nucleons  (where $\gamma_I\overline{{\bm I}}_i$ is the effective magnetic moment considering the electric quadrupole interaction for ${\bm I}_i\ge1$).

On the other hand, if the coordination of the nuclear magnetic moment viewed from the muon is isotropic and the number of coordination is sufficiently large ($\ge4$), the classical spin treatment is easier, and the density distribution $n({\bm H})$ of ${\bm H}({\bm r})$ is approximated by a Gaussian distribution with zero mean value,
\begin{eqnarray}
n(H_\alpha)&=&\langle\delta(H_\alpha-H_\alpha({\bm r}))\rangle_{\bm r}\nonumber
\\&=&\frac{\gamma_{\mu}}{\sqrt{2\pi}\Delta}
\exp\left(-\frac{\gamma_{\mu}^{2}H_\alpha^{2}}{2\Delta^{2}}\right)  \:\:   (\alpha=x,y,z)
\label{ph}
\end{eqnarray}
Here, $\Delta$ is given by the second moment of ${\bm H}_{\rm d}$ as
\begin{equation}
\frac{\Delta^2}{\gamma_\mu^2}=\sum_i\sum_{\alpha,\beta}[\gamma_I(\hat{A}_{\rm d}^i)^{\alpha\beta}\overline{{\bm I}}_i]^2, \label{delta_n}
\end{equation}
with $\beta$ taking all $x,y,z$, and the $\alpha$ over the $x,y$ components that are effective for longitudinal relaxation when $\hat{z}$ is the longitudinal direction; the $z$ component does not contribute to the relaxation because it gives a magnetic field parallel to the muon spin.
In this case, the spin polarization ${\bm G}(t)$ is given by the motion of one muon spin projected onto ${\bm H}$ with the angle between the magnetic field ${\bm H}$ and the $\hat{z}$ axis as $\theta$,
\begin{equation}
P_z(t)=\cos^2\theta+\sin^2\theta\cos(\gamma_\mu Ht)\label{sz}
\end{equation}
which is averaged by $n({\bm H})$ in Eq.~(\ref{ph}) to yield the Kubo-Toyabe function
\begin{eqnarray}
G_z(t)&=&\iiint_{-\infty}^{\infty}P_{z}(t)\Pi_\alpha n_\alpha(H_{\alpha})dH_\alpha\nonumber\\
&=&\frac{1}{3}+\frac{2}{3}(1-\Delta^{2}t^{2})e^{-\frac{1}{2}\Delta^{2}t^{2}}. \label{gkt}
\end{eqnarray}
The magnitude of $\Delta$ is sensitive to the size of the nearest-neighbor nuclear magnetic moment $\gamma_I\overline{{\bm I}}_i$ and the distance $r_i$ from the muon, and the position occupied by the muon as pseudo-hydrogen can be estimated by comparing the experimentally obtained $\Delta$ with the calculated value at the candidate sites. Especially in recent years, the reliability of the first-principles calculations based on density functional theory (DFT) have been improved, and by using this method to narrow down the candidate sites, the muon sites can be estimated with higher credibility.

\section*{APPENDIX B: \msr\ spectrum in the presence of unpaired electrons}
As in the previous section, the host is assumed to be a nonmagnetic material. In this case, the unpaired electron originates from that bound to the muon (Mu$^0$) (corresponding to Fig. \ref{muc}c, d, e, and g). In general, the Hamiltonian for the magnetic interaction between muon and unpaired electron is given by
\begin{eqnarray}
{\cal H}/\hbar &=& [{\cal H}_{\rm d}+{\cal H}_e+{\cal H}_{\rm Mu}]/\hbar,\label{Htot}\\
{\cal H}_{\rm Mu}/\hbar&=&\gamma_\mu\gamma_e{\bm S}_e[\frac{8\pi}{3}\delta({\bm r})+\hat{A}_{\rm d}]{\bm S}_\mu\label{HMu}\\
& = &\frac{1}{2}(2\pi{\bm A})\cdot{\bm S}_\mu=\frac{1}{2}{\bm \omega}_{\rm Mu}\cdot{\bm S}_\mu, \label{Hspin}
\end{eqnarray}
where ${\cal H}_{\rm d}$ is the muon-nuclear spin system [Eq.~(\ref{nucldip})], and ${\cal H}_e$ is the Hamiltonian of the electron system with $\gamma_e$ being the gyromagnetic ratio of the electron ($=2\pi\times28.02421$ GHz/T).  The first term in ${\cal H}_{\rm Mu}$ is for the Fermi contact interaction, and the second term is for the magnetic dipolar interaction. 

Provided that the magnitudes of interactions between nuclear spins and muons/electrons are negligible, Eq.~(\ref{Htot}) is a two-spin Hamiltonian whose eigenstates are given by the linear combination of the muon electron spin eigenfunctions $|s_z^\mu,s_z^e\rangle$ ($s_z^\mu,s_z^e=\pm 1/2$), with four corresponding eigenenergies ($E_m$, $m=1$--4).
When an external magnetic field ${\bm H}_0$ is applied, the spin rotation signals corresponding to the allowed transitions between these eigenstates,
\begin{equation}
G(t)=\sum_{n<m}a_{nm}\cos\omega_{nm}t,\label{pmux}
\end{equation}
are observed, where $\omega_{nm}=\omega_n-\omega_m=(E_n-E_m)/\hbar$ are the spin rotation frequencies and $a_{nm}$ are their amplitudes.

Now, taking $\hat{z}$ as the main axis of the hyperfine interaction ${\bm \omega}_{\rm Mu}$ with $\theta$ and $\phi$ being the polar and the azimuthal angles, Eq.~(\ref{Hspin}) is expressed as
\begin{eqnarray}
{\cal H}_{\rm Mu}/\hbar&=&\frac{1}{2}\omega_{\rm Mu}(\theta,\phi) = \frac{1}{2}2\pi A(\theta,\phi),\nonumber\\
A(\theta,\phi)&=&(A_x\cos^2\phi +A_y\sin^2\phi)\sin^2\theta \nonumber\\
& &+A_z\cos^2\theta, \label{Amu}
\end{eqnarray}with which we can sort out the qualitative relationship between the electronic structure of Mu$^0$ with surrounding atoms and that of $A(\theta,\phi)$.

The reason for the formation of bound states is the relatively weak local dielectric shielding (determined by the permittivity $\epsilon$) that leads to the long-range Coulomb interaction between muons and electrons. If the bound electron is in a 1$s$ orbital-like state (as shown in Fig.~\ref{muc}c), the hyperfine interaction is dominated by the Fermi contact term and is isotropic with positive sign as a whole ($A_x\simeq A_y\simeq A_z>0$). In this case, the absolute value of $A$, the effective Bohr radius, and the depth of the bound level are estimated to be 
\begin{eqnarray}
A&\simeq& \frac{\omega_{\rm vac}}{2\pi}\left[\frac{m^*}{\epsilon' m_e}\right]^3,\label{efm}\\
a^*&\simeq& a_0  \frac{\epsilon' m^*}{m_e},\\
R^*&\simeq& R_y\frac{m^*}{\epsilon^{'2}m_e},
\end{eqnarray}
where $m^*$ is the effective mass of the electron in the conduction band, $R_y$ is Rydberg's constant, and $\epsilon'$ is the relative permittivity at zero frequency [$=\epsilon(\omega\rightarrow0)/\epsilon$]. This is thought to be one of the mechanisms by which shallow donor levels are induced by interstitial hydrogen in semiconductors with high permittivity.

However, such a Jellium model is not sufficient for actual materials, and the electronic states of H/Mu are anisotropically distributed due to interactions with surrounding atoms. One such example is Mu$^0$ located near the bonding center between host atoms, which has been known for a long time in elemental semiconductors with diamond structure and in group 13-15 compound semiconductors such as GaAs with zinc blende structure (corresponding to Mu$_D$ in Fig.~\ref{mue}b) \cite{Patterson:88}.  In these examples, the hosts have a four-coordinate configuration with $sp^3$ hybrid orbitals that are strongly covalent, and Mu/H breaks this bond to make a new bond with the anion ($I^-$), and the excess electrons become a dangling bond on the cation ($K^+$) (see Fig.~\ref{mue}c,d).  In this case, the hyperfine interaction has an anisotropy symmetric around the axis connecting the muon and the electron, and by taking the symmetry axis to $\hat{z}$,  Eq.~(\ref{Amu}) is reduced to
\begin{equation}
A(\theta,\phi)=A^*+D\cos^2\theta \label{Amu2}
\end{equation}
with $A_{x,y}\equiv A_\perp\equiv A^*-D/2$ and $A_z\equiv A_\parallel\equiv A^*+D$.
The first reported Mu$^0$ with shallow electronic levels were found in II-VI compounds such as zinc oxide (ZnO) \cite{Cox:01,Shimomura:02}, which also exhibits hyperfine interactions well described by Eq.~(\ref{Amu2}).

\section*{APPENDIX C: A model for the H/Mu site occupancy of the asymmetric double-well potential}
We consider a simple model for the temperature dependence of the H/Mu site occupancy in the presence of two sites (Site-$A$ and -$D$, see Fig.~\ref{adia}a) with asymmetric double-well potential separated by a potential barrier ($V$ and $V'$). Provided that $V'\gg V$, the partition function for the two-level system is approximately given by
\begin{align}
  Z(\beta)=n_D+n_Ae^{-\beta V},
\end{align}
where $\beta\equiv 1/k_BT$, $n_A$ and $n_D$ are the degeneracy of the each site in the unit cell. The fractional occupancy of Mu/H for the respective sites in the equilibrium state is then described by
\begin{align}
  f_A=\frac{n_Ae^{-\beta V}}{n_D+n_Ae^{-\beta V}},\label{fv}\\
  f_D=\frac{n_D}{n_D+n_Ae^{-\beta V}}\label{fvv}.
\end{align}
Note that $f_D<1$ at finite temperatures ($n_Ae^{-\beta V}>0$) to reduce the free energy by gaining entropy.

Now, we presume that the initial site occupancy for Mu is that quenched from $T=\infty$ ($\beta=0$), so that $f_A=n_A/(n_A+n_D)\equiv f^0_A$, $f_D=n_D/(n_A+n_D)\equiv f^0_D$ (i.e., proportional to the number density of available sites). Then, the fractional yields observed by $\mu$SR at the finite temperature correspond to the average fraction of muons over the annealing process from this initial distribution to the thermal equilibrium distribution (Eqs.~\ref{fv},~\ref{fvv}) in the time scale of $\sim$$10^1$ $\mu$s. Such a relaxation process is generally described by the fluctuation-dissipation theorem within the linear response theory for the macroscopic systems.

\begin{figure}[t]
  \centering
  \includegraphics[width=0.9\linewidth]{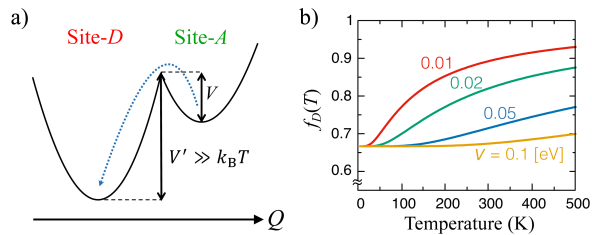}
  \caption{(a): Schematic drawing of the two-site model based on the Boltzmann factor. (b): Temperature dependence of Eq.~(\ref{Eq_twosite}) for various $V$  with $f_D^0=2/3$, $f_A^0=1/3$).}
  \label{twosite}
\end{figure}

However, since the implanted Mu as microscopic entity probes the local fluctuations only, we assume that the observed temperature dependence of $f_D$ is determined by the migration from Site-$A$ to  Site-$D$ via thermally activated hopping over a potential barrier $V$ [see Fig.~\ref{twosite}~(a)], where the migration probability is proportional to $e^{-\beta V}$. The observed fraction of muons at Site-$A$ can be approximately given by $f_A(T)\simeq f_A^0\left(1-e^{-\beta V}\right)$, which is valid for low temperatures ($V'\gg k_BT$) where the inverse hopping process is negligible. Thus have
\begin{align}
  f_D(T)&=1-f_A(T)\nonumber \\
    &\simeq f_D^0+f_A^0e^{-\beta V},
  \label{Eq_twosite}
\end{align}
for the temperature dependence of the Mu occupancy at Site-$D$.
Fig.~\ref{twosite}~(b) shows examples of $f_D(T)$ given by Eq.~(\ref{Eq_twosite}) for various $V$.

%

\end{document}